\documentclass[aps,showpacs,pra,amssymb, amsmath, twocolumn]{revtex4}
\usepackage{graphicx}
\usepackage{amssymb}
\usepackage{amsmath}
\usepackage{bm}
\usepackage{xcolor} 
\usepackage{array}
\usepackage{multirow}
\usepackage{tabularx}
\usepackage{booktabs}
\usepackage{textcomp}
\usepackage{color,soul}
\usepackage{microtype}

\begin{document}
\title{Machine learning identification of symmetrized base states of Rydberg atoms}
\author{Daryl Ryan Chong$^1$, Minhyuk Kim$^2$, Jaewook Ahn$^2$, and Heejeong Jeong$^1$}
\email{jhj413@gmail.com}
\address{$^1$Department of Physics, Faculty of Science, University of Malaya, Kuala Lumpur 50603, Malaysia}
\address{$^2$Department of Physics, KAIST, Daejeon 34141, Korea}

\begin{abstract}
\noindent
Studying the complex quantum dynamics of interacting many-body systems is one of the most challenging areas in modern physics. Here, we use machine learning (ML) models to identify the symmetrized base states of interacting Rydberg atoms of various atom numbers (up to six) and geometric configurations. To obtain the data set for training the ML classiﬁers, we generate Rydberg excitation probability proﬁles that simulate experimental data by utilizing Lindblad equations that incorporate laser intensities and phase noise. Then, we classify the data sets using support vector machines (SVMs) and random forest classiﬁers (RFCs). With these ML models, we achieve high accuracy of up to 100\% for data sets containing only a few hundred samples, especially for the closed atom conﬁgurations such as the pentagonal (ﬁve atoms) and hexagonal (six atoms) systems. The results demonstrate that computationally cost-effective ML models can be used in the identification of Rydberg atom configurations.
\end{abstract} 
\maketitle
\section{Introduction: Background and Motivation}
\label{intro}
Rydberg-atom quantum simulators currently draw much attention due to the scalability and diverse multi-qubit conﬁgurations of single-atom trap arrays. Three dimensional (3D) 100 Rydberg atom-arrays \cite{Barredo2018} and fast processing of holographic 3D arrays \cite{Sun2021} have been recently demonstrated. However, as the number of atoms increases, the quantum dynamics of interacting Rydberg atoms become significantly more complex compared to few-qubit Rydberg atom cases. In real situations of nonzero decoherence, the resulting partially-entangled $N$-body states, in general, require data processing of $2^N$-by-$2^N$ density matrices, which is a formidable task for large $N$. Hamiltonian symmetries shall reduce the computational complexity and identiﬁcation of the symmetrized base states of the interacting Rydberg atoms shall be of great benefit for further studies of Rydberg quantum simulators.

This paper intends to use machine learning (ML) methods to identify the symmetrized base states of interacting many-body systems. ML has started to affect many areas of our everyday life in recent years, from personalized internet search algorithms and voice recognition to diverse decision-making activities. ML is used in physics research \cite{Carleo2019} to predict the bandgaps of inorganic compounds \cite{Lee2016}, represent topological ground states as short-range neural networks \cite{Deng2017}, and optimize the electrochemical properties of battery materials \cite{Min2018}. In quantum physics research, ML has been implemented to reconstruct the density matrices in quantum state tomography \cite{Tolai2018} and quantum motional state tomography for levitated particles \cite{Weiss2019}, retrieve topological quantum phase transitions \cite{Che2020}, and also perform optimum parameter search \cite{Barker2020}. Such needs become popular in the broad range of quantum information science as the parametric space (e.g. qubit numbers) increases, thereby resulting in greater complexities.

Most ML models implemented in quantum research up until now have utilized artiﬁcial neural networks (ANNs), which require large amounts of computational resources, data sets with large sample sizes, and longer training time. However, these general requirements of ANNs may not be fully necessary for certain tasks. As to be shown below, our Rydberg-identiﬁer (Rydberg-ID) of intrinsic symmetries is a good example. To the best of our knowledge, ML models have not been used for classification tasks in Rydberg-atom configurations.

\begin{figure*}[tbp!]
  \centering
\includegraphics[width=1.0\textwidth]{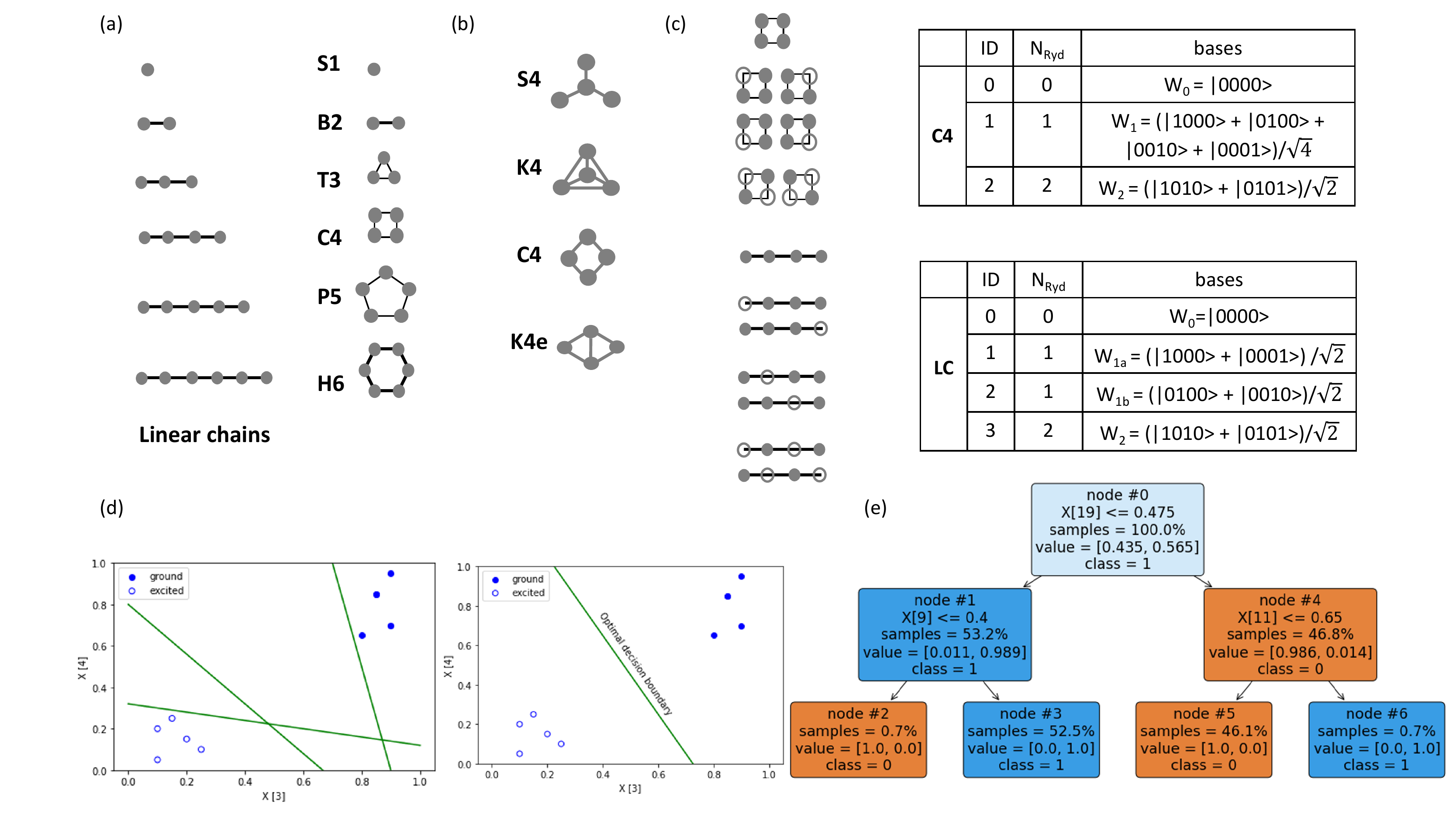}

\caption{Schematics of considered Rydberg-atom configurations. Classification by (a) the number of atoms from one to six, (b) various connected graphs of 4 atoms, and (c) Rydberg excitation. (d) Illustration of the possible decision boundaries to classify the ground state and excited single atoms and SVM's optimal decision boundary. (e) A decision tree from the random forest classifier used to classify the ground state and excited single atoms generated with 
\textbf{plot\_tree()}, as explained in section \ref{RFC}
}
\label{Fig1}
\end{figure*}

\section{Research Methods: Collecting Rydberg Quantum Simulation Data}
\label{matlab}
We simulate experimental data sets of temporal Rydberg excitation probabilities to train and test the identiﬁcation of its atom conﬁguration (i.e. number and shape). Each configuration corresponds to its interaction Hamiltonian, the second term $U(r_{jk})$ in Eq. (\ref{eq:Hamiltonian}) below if we consider Rydberg fraction. For the classification, we assign an identifier ($ID = 0,1,2,...$) to each Rydberg base state from the interaction Hamiltonian, $U(r_{jk})$. Here, we use two supervised machine learning algorithms implemented with a python machine learning library: scikit-learn \cite{scikit-learn}. Support vector machine (SVM) and random forest classifier (RFC) models are developed for the Rydberg-ID. Both models show accuracy up to 100\% within the 2\% error range with only a few hundred data samples for each atomic configuration up to four Rydberg atoms. The classifications consistently show high accuracy, close to 100\% for all the closed atomic configurations up to six atoms. We define the following variables of $N$ related to certain numbers used: \\
$N$: Number of samples in a data set \\
$N_{a}$: Number of atoms, which is qubit numbers.\\
$N_{Ryd}$: Number of atoms in the Rydberg excited state.\\
$N_{ID}$: Number of possible quantum base states, $W_{N_{Ryd}}^{G}$, where the superscript $G$ denotes a specific graph. \\

Considering the Rydberg-atom blockade regime \cite{Lukin2001}, we use various connected graphs consisting of vertices and edges that represent atoms and blockaded couplings \cite{Kim2020} respectively, as illustrated in Fig. \ref{Fig1} (a)-(c). Each configuration corresponds to Rydberg base states of a Rydberg interaction Hamiltonian that generates time-evolution probability. In general, we can write the $N_a$ atom Rydberg Hamiltonian as,

\begin{eqnarray}
\hat{H}=\frac{\hbar\Omega}{2} \sum_{j=1}^{N_a} (|1\rangle_j\langle0|_j+|0\rangle_j\langle1|_j)+\sum_{j<k}U(r_{jk})\hat{n}_j \hat{n}_k,\label{eq:Hamiltonian}
\end{eqnarray}
where $|0\rangle_j$ and $|1\rangle_j$ are the ground and Rydberg energy states of an atom $j$ located at $\Vec{r}_j$ respectively. The second term $U(r_{jk})\equiv C_6/|\Vec{r}_{jk}|^6$ corresponds to the van der Waals interaction between two Rydberg atoms, where $\Vec{r}_{jk} \equiv \Vec{r}_j-\Vec{r}_k$. $\hat{n}_j=|1\rangle_j\langle1|_j$ is the excitation number \cite{Lukin2001, Kim2020}. In the graphs representing atomic arrangements, we set the length of the edges to be $d = 8 ~\mu$m, which is within the range of the Rydberg-blockade radius $d < r_b \equiv |C_6/\hbar\Omega|^{1/6}$. Thus, in this case, Rydberg excitation of an atom only occurs whenever its neighboring atoms are in the ground state. The distance $\Vec{r}_{jk}$ differs according to the geometry of the atomic arrangements. For example, $\Vec{r}_{jk}=2 d$ for atoms at opposite vertices in the hexagonal configuration (H6). To simulate real experimental data, we add the next nearest neighbor interaction (NNN) between two Rydberg atoms $U(r_{jk})$. 
For linear chains, we may consider only the nearest neighbor (NN) interaction between atoms. However, considering actual experimental data, it is more accurate to add the next nearest neighbor interaction. Adding the next nearest neighbor interaction increases the accuracy significantly for linear chains of four and five atoms, as discussed later in Fig. \ref{Fig7}.

We model our ML classification based on the experiments in Ref \cite{Kim2020}. The experimental process starts with the Rubidium ($^{87}$Rb) magneto-optical trap (MOT) and single atom arrays in three-dimensional optical tweezers via dipole trap laser. The atom arrangements are determined by a phase mask and a spatial light modulator (SLM) that control the focal points of the dipole trap laser. All atoms are optically pumped to the ground hyperfine state $|0\rangle=|5S_{1/2} , F = 2, m_F = 2\rangle$ as initial states. We then implement the Hamiltonian of Eq. (\ref{eq:Hamiltonian}) by the two-photon excitation (780-nm and 480-nm lasers) of the Rydberg atoms to $|1\rangle=|71S_{1/2}, m_J = 1/2\rangle$ via the off-resonant intermediate state $|m\rangle=|5P_{3/2}, F' = 3, m'_F = 3 \rangle$. By the final state readout process \cite{Kim2020}, we may obtain the experimental probability data to which we can apply our ML classifiers. We implement the two Rydberg laser noises in our simulation to generate the  experimental-like data used in our ML models.

For generating experimental-like data, we include decoherence by implementing the Lindblad equation in our simulation using MATLAB. In detail, we set the dephasing rates associated with the laser noise, including intensity fluctuation. The phase noise of lasers is measured and imported. 
To consider the situation of dephasing fluctuations, we randomly generate the intensity fluctuation by setting the mean value of the fluctuation (3\%) and its deviation (1\%). 
Here, we consider the typical excitation scheme to Rydberg states via two-photon transition \cite{Urban2009,Gaetan2009}, so the Lindblad model is included to consider the dissipation to the intermediate transition level. 
We randomly generate the intensity fluctuation by setting the mean value of the fluctuation (3\%) and its deviation (1\%). Consequently, the Lindblad model's dephasing rates fluctuate due to the laser intensity-induced power broadening.

We compared classification accuracy for other mean values (deviations) of the laser intensity fluctuation for three combined sets - 3\% (1\%), 7\% (2\%), and 10\% (5\%), as we discuss later in Fig.\ref{Fig14} and Fig.\ref{Fig15}. Both 3\% (1\%) and 7\% (2\%) represent realistic experimental situations, and the accuracy does not change much for these two. However, considerable fluctuation with 10\% (5\%) that lowers the accuracy does not represent our experimental condition.

For $N_a$ atom qubit configuration, only a few out of the $2^{N_a}$ eigenstates represent the possible Rydberg-atom excitation (bright eigenstates) and its ground state. Therefore, we only consider possible excitations as our configurational basis. Time evolution of the state, $|\Psi(t)\rangle$, from the initial ground state $W_0 \equiv |\Psi(t=0)\rangle = |00...0\rangle$ [Rydberg-atom base state, $W_{N_{Ryd}} \equiv |\Psi_{N_{Ryd}}\rangle$], is determined by the Hamiltonian [Eq.(\ref{eq:Hamiltonian})] in the Lindblad model as discussed in Ref \cite{Kim2020},

\begin{eqnarray}
P_0(t) = |\langle \Psi(0)|\Psi(t)\rangle|^2, \label{eq:prob_ground}\\
P_{N_{Ryd}}(t) = |\langle \Psi_{N_{Ryd}}|\Psi(t)\rangle|^2, \label{eq:prob_excited}
\end{eqnarray}
where $P_0(t)$ [$P_{N_{Ryd}}(t)$] is the probability amplitude at time $t$. The time evolution becomes complicated and negligible as we increase the number of atoms $N_a$. From the data of certain Rydberg base states that do not show much distinction, we may ask questions such as:\\ 

\noindent 1. Can we count the number of atoms, $N_a$, in the ground state? [Fig.\ref{Fig1} (a)]\\

\noindent 2. With a fixed number of atoms, can we identify different geometric configurations,  $W_{0}^{G}$, in their ground state \cite{Kim2020}? [Fig.\ref{Fig1} (b)]\\

\noindent 3. With fixed configuration, can we classify all possible Rydberg quantum base states, $W_{N_{Ryd}}^{G}$? [Fig.\ref{Fig1} (c)]\\

The three questions above can be considered as a reverse engineering process if we can determine the Hamiltonian or base state of the system from its probability data, $P_0(t) $ or $P_{N_{Ryd}}(t)$. The motivation of this paper is to perform such classification (Rydberg-ID) tasks using machine learning models.

\section{Research Methods: Machine Learning Algorithms}

In this section, we briefly introduce two supervised classification models that we implement as our Rydberg-ID: support vector machines (SVMs) and random forest classifiers (RFCs). Both methods are built with scikit-learn \cite{HOML}.

The method described in section \ref{matlab} generated the Rydberg atoms' temporal probabilities of being in a certain state for the ML data sets. Each probability plot is collected for a duration of 1 $\mu$s. As is the norm for real experiments, each probability data point is generated every 50 ns. Thus, each plot has 21 data points. The notation $X [i]$ represents a probability value at the $i^{th}$ data point representing a certain time, $t$. These data points are used as the features for our ML models.
The term \emph{feature} means a variable that describes the samples used to train the ML model. We assume that the probability plots of the configurations are sufficiently different from each other such that the models can classify them.

\subsection{Support Vector Machines}

Figure \ref{Fig1} (d) illustrates the main idea of support vector machines (SVMs) with a few samples from the single-atom data set in Fig \ref{Fig2} (a) and (c). The axes denote the probability value at the 3$^{rd}$ and 4$^{th}$ time data points; only 2 out of the 21 features are plotted for visualization. The green lines are \emph{decision boundaries} that classify the data points into the two classes: ground state and excited state. 

An SVM determines the optimal decision boundary that separates the classes and maximizes the distance between the decision boundary and the nearest data point from each class, as shown on the right side of  Fig \ref{Fig1} (d). For data with $p$ features, the decision boundary is a ($p-1$)-dimensional hyperplane. We build the linear SVM model with scikit-learn's \textbf{SGDClassifier()} with the default loss function: ``hinge''. In the model, we scaled the training data with \textbf{StandardScaler()} to improve the performance of the SVM since SVMs are sensitive to the scales of the features \cite{HOML}.

\subsection{Random forests}
\label{RFC}
Contrary to the SVM, a random forest is an \emph{ensemble method} that has more than one predictor and combines the predictions of the predictors to improve its performance \cite{HOML}. The final prediction of the ensemble method is the average of all the individual predictions. 
An ensemble method usually has better performance than its constituent predictors, especially if the constituent predictors make diverse errors. Random forests consist of \emph{decision trees} diversified by being trained on random subsets of the samples and features.

Figure \ref{Fig1} (e) shows a decision tree from the random forest classifier used in Fig.\ref{Fig2} (c) to classify the single-atom data set. The decision tree learns the rules at each node using the Classification and Regression Tree (CART) algorithm \cite{HOML} with randomly selected training samples. If a sample fulfills the rule at a decision tree node, it moves down to the left and vice versa. In each node, \emph{sample} represents the percentage of training samples that pass through by satisfying the condition in its parent node. The \emph{value} shows the proportion of the training samples belonging to each class determined by its known ID. The \emph{class} denotes the predicted class of the samples, which is the node's mode value. Similar to the SVM model, we build a random forest classifier (RFC) with scikit learn's \textbf{RandomForestClassifier()}. We used the default parameters of the models provided by scikit-learn \cite{scikit-learn}. 

\subsection{Training and Testing of Models}

To train and test the ML models, we generated 300 data samples for each atomic configuration. Consequently, the total number of data samples is $N =300 \times N_{ID}$, where $N_{ID}$ is the total number of atomic configurations considered in the experiment.
For each data set, we split it into a training set with 80\% of the samples and a test set with the remaining 20\% of samples. The training sets are used to build the classification model, and the test sets are used to estimate the generalization error. To create the training and test sets, we use stratified sampling with scikit-learn's \textbf{train\_test\_split()} function to ensure our training and test sets have equal proportions of $N_{ID}$ similar to the whole data set. We label the samples in the data sets with an identifier (ID) from zero to $N_{ID}-1$ as shown in the figures for these supervised learning tasks.

We test the models on the test set and obtain the models' estimated accuracy on other data that we did not use in training before. The accuracy is simply the proportion of correct predictions of each model. To quantify the precision of accuracy on the test set, we calculated the 95\% confidence interval. The correct interpretation of the confidence interval can be achieved as follows. If we had more data sets and created their confidence intervals, almost 95\% of the constructed confidence intervals would contain the true classifier's accuracy \cite{Cox1974}. One may misinterpret that there is a 95\% probability that the true classifier's accuracy is within the confidence interval \cite{Hoekstra2014}. 
We view each prediction as a Bernoulli trial with the probability of success equal to the model's accuracy. Thus, we create a binomial proportion confidence interval and use the Agresti–Coull interval as it is preferred for large sample sizes ($N>40$) \cite{Brown2001}.


\section{Results}

\begin{figure*}[tbp!]
  \centering
\includegraphics[width=1.0\textwidth]{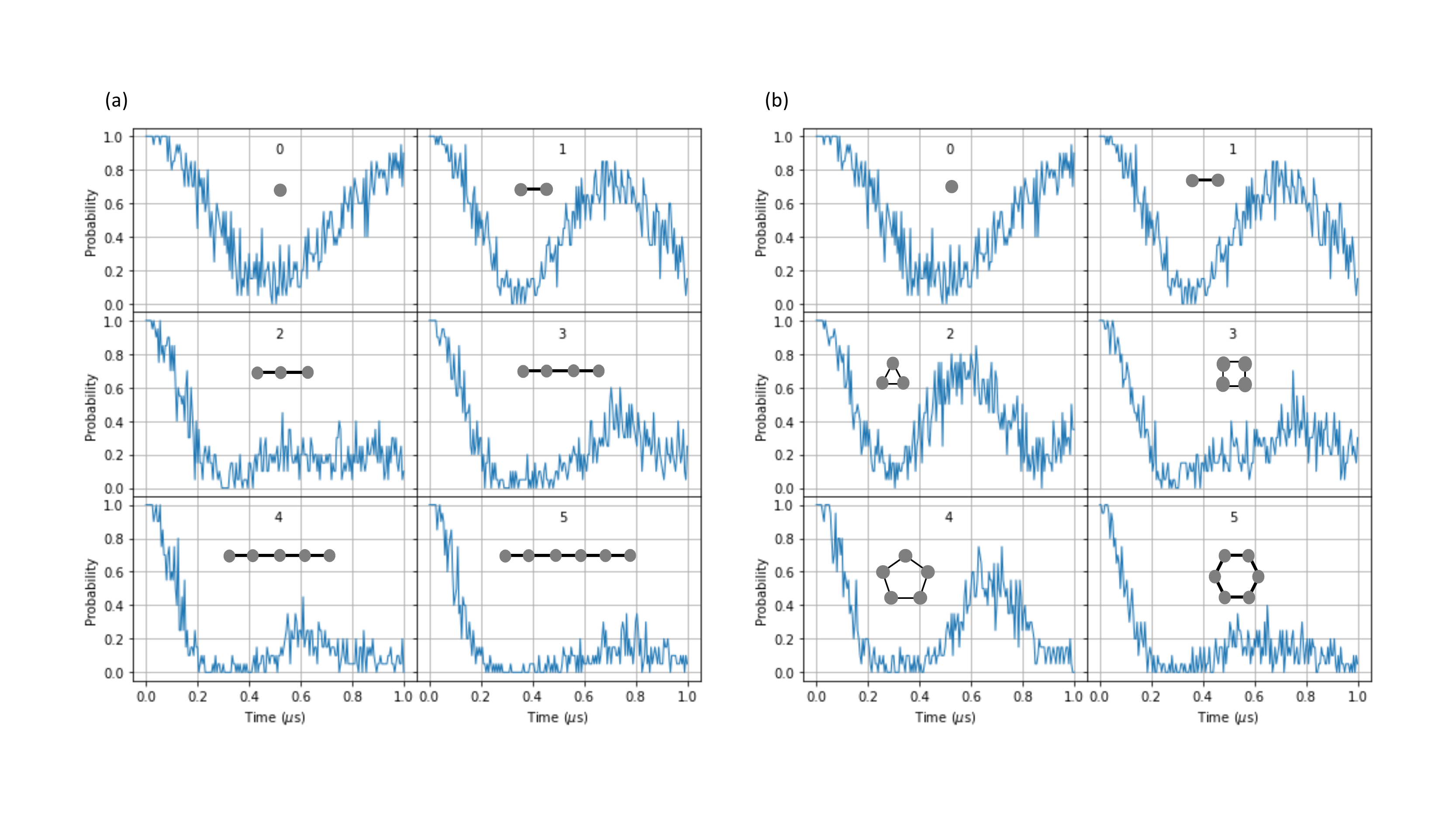}\\
\includegraphics[width=1.0\textwidth]{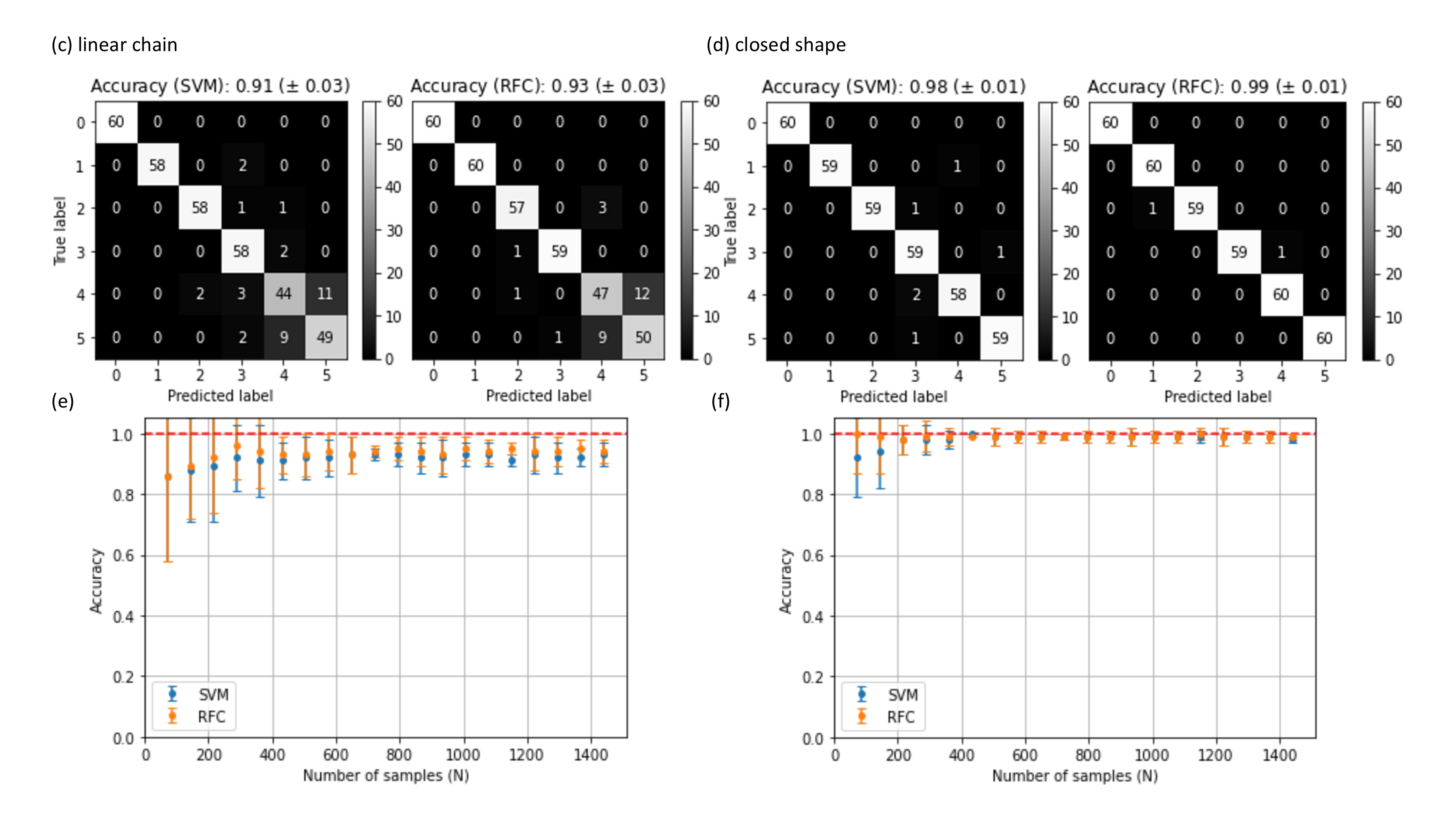}
\caption{Identifying the number of atoms in its Rydberg ground states: $W_0 \equiv |\Psi(0)\rangle=|0>$, $|$00$>$, $|$000$>$, $|$0000$>$, $|$00000$>$, $|$000000$>$, from one to six ground state atoms either in linear chains (a,c,d)  or closed shapes (b,d,f), respectively. We plot time evolution of the probability $P_0(t) = |\langle \Psi(0)|\Psi(t)\rangle|^2$ [Eq. (\ref{eq:prob_ground})] in (a) and (b). We plot the confusion matrix for the (c) linear chains, and (d) closed shape using two ML methods: SVM and RFC. The accuracy of the test sets is shown in (e) and (f) for both SVM (blue dots) and RFC (orange dots) methods, as a function of sample numbers $N$. The saturated accuracy of the linear chains is 91 $\sim$ 93 \%, which is less than the 100 \% accuracy of the closed shape. The reduced accuracy results from the off-diagonal elements of the confusion matrix in (c), which shows the mixture between $ID =$ 4 and 5 corresponding to their similarity in probabilities $P_0(t)$. }
\label{Fig8}
\end{figure*}

\begin{figure*}[tbp!]
  \centering
\includegraphics[width=0.85\textwidth]{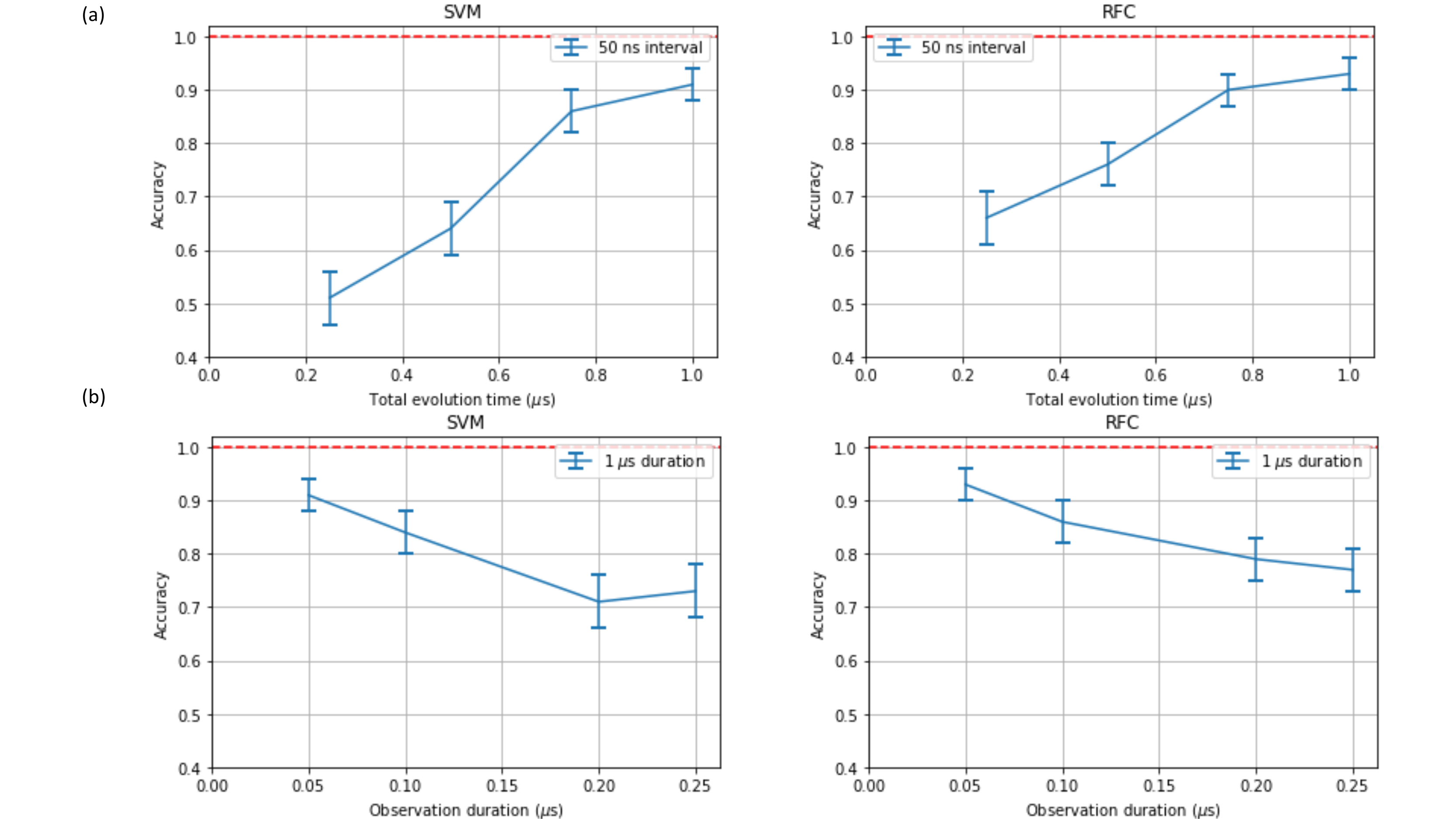}
\caption{Accuracy performance depending on (a) total evolution time, and (b) observation duration between data points for the case of identification of $N_a$ in Fig.~\ref{Fig8} (a,c,e). The accuracy values are obtained by utilizing both SVM and RFC. The default total evolution time is 1 $\mu$s, and the default duration is 50 ns for all our ML sample data in this paper. The results show our default experimental parameters provides above 90 \% accuracy as in Fig.~\ref{Fig8} (a,c,e).}
\label{FigRev}
\end{figure*}

\subsection{Classification by the number of atoms up to six}

First, we identify the number of atoms in the ground state by considering two types of configurations: linear chains and closed shapes. We increase the number of atoms in line from one to six (ID: 0, 1, 2, 3, 4, 5). Figure \ref{Fig8} (a), shows time evolution of the probability $P_0(t) = |\langle \Psi(0)|\Psi(t)\rangle|^2$ in Eq. (\ref{eq:prob_ground}) for each case (ID). The models are trained on temporal probability plots with 21 data points corresponding to the 50 ns duration between each data point similar to our model experiment in Ref \cite{Kim2020}. However, the 21 points are not enough to show all features of the temporal profiles. Thus, for better visualization in the figures, we simulate a single temporal profile with 201 data points for each Rydberg configuration set.

To check classification performances, we tested the accuracy depending on different 1) total evolution time and 2) observation duration in Fig. \ref{FigRev}. The accuracy values are compared to Fig. \ref{Fig8} (a,c,e) obtained by our default parameter settings, which is 1 $\mu$s total evolution time and 50 ns duration. For the identification of the number of atoms $N_a$, our accuracy tests employ both SVM and RFC. Fig. \ref{FigRev} (a) shows that the SVM [RFC] accuracy drops from 91\% ($\pm 0.03$) [93\% ($\pm 0.03$)] to 86\% ($\pm 0.04$) [90\% ($\pm 0.03$)] when we decrease the overall time evolution from 1 $\mu$s to 0.75 $\mu$s while keeping the observation duration constant, 50 ns. It drops further to 51 \% ($\pm 0.05$) [66 \% ($\pm 0.05$)] at 0.25 $\mu$s.

Fig. \ref{FigRev} (b) shows that the SVM [RFC] accuracy drops as we increase the observation duration from 50 ns to 250 ns while keeping the overall time evolution constant, 1 $\mu$s. The overall performance is higher with RFC compared to SVM, although both show similar trends. The results confirm that the classification performance decreases as the range of evolution time decreases and the duration increases. The reason is that reducing the total evolution time and increasing the observation duration lessens the amount of information present in the data set, which hampers the performance of the ML models. This is because ML models usually perform better when they are trained on more data, provided that the data is reliable. Note that our default setting maintains the overall accuracy above 90\%.

Figure \ref{Fig8} (c) shows the accuracy of the models on the test set rendering the $N_{ID} \times N_{ID}$ confusion matrix for both SVM and RFC. We use the confusion matrix to assess ML models as it shows the predicted ID of the samples together with its actual ID; the diagonal elements represent correct predictions while the off-diagonal elements indicate the wrong ones. For example, non-negligible off-diagonal elements between ID $=$ 4 and 5 indicate incorrect identification between five and six atomic linear chains. Then, we need to maximize the diagonal values representing ``true positive (TP)'' predictions.

To estimate the minimum number of samples required to train our models, we performed 10-fold cross-validation on the training sets while varying their sample size. Figure \ref{Fig8} (e) plots the cross-validated accuracy of the training set as a function of N.  The accuracy of the linear-chain data set [Fig. \ref{Fig8} (e)] is less than the closed-shape data set [Fig. \ref{Fig8} (f)] because of the similarity between probability data shapes (ID $=$ 4 and 5) as shown in Fig \ref{Fig8} (a). The similarity of the data of different classes affects the prediction accuracy. If the data of different classes are linearly separable in the feature space, as in Fig 1 (d), the accuracy of the ML models would be higher. However, if the data of different classes are very close to one another, the accuracy of the SVM and RFC would be lower. Such similar patterns also explain the reason for the significant off-diagonal elements between five and six linear chains. 

Contrary to the linear chains with open ends, we define the following closed shapes for the same number of atoms. There is no such closed shape for single and two (binary) atoms, S1 and B2. The closed two-dimensional shapes begin with three-atom configurations: triangular (T3), square (C4), pentagonal (P5), and hexagonal (H6) configurations, as shown in Fig.\ref{Fig1} (a). Figure \ref{Fig8} (b) shows the probabilities of being in the base state for each closed shape. We obtain higher accuracy of 98\% from SVM and 99\% from RFC [Fig.\ref{Fig8} (d)]. The saturation accuracy of 99\% occurs near 200 samples (Fig.\ref{Fig8} (f)). 

\begin{figure*}[tbp!]
  \centering
\includegraphics[width=1.0\textwidth]{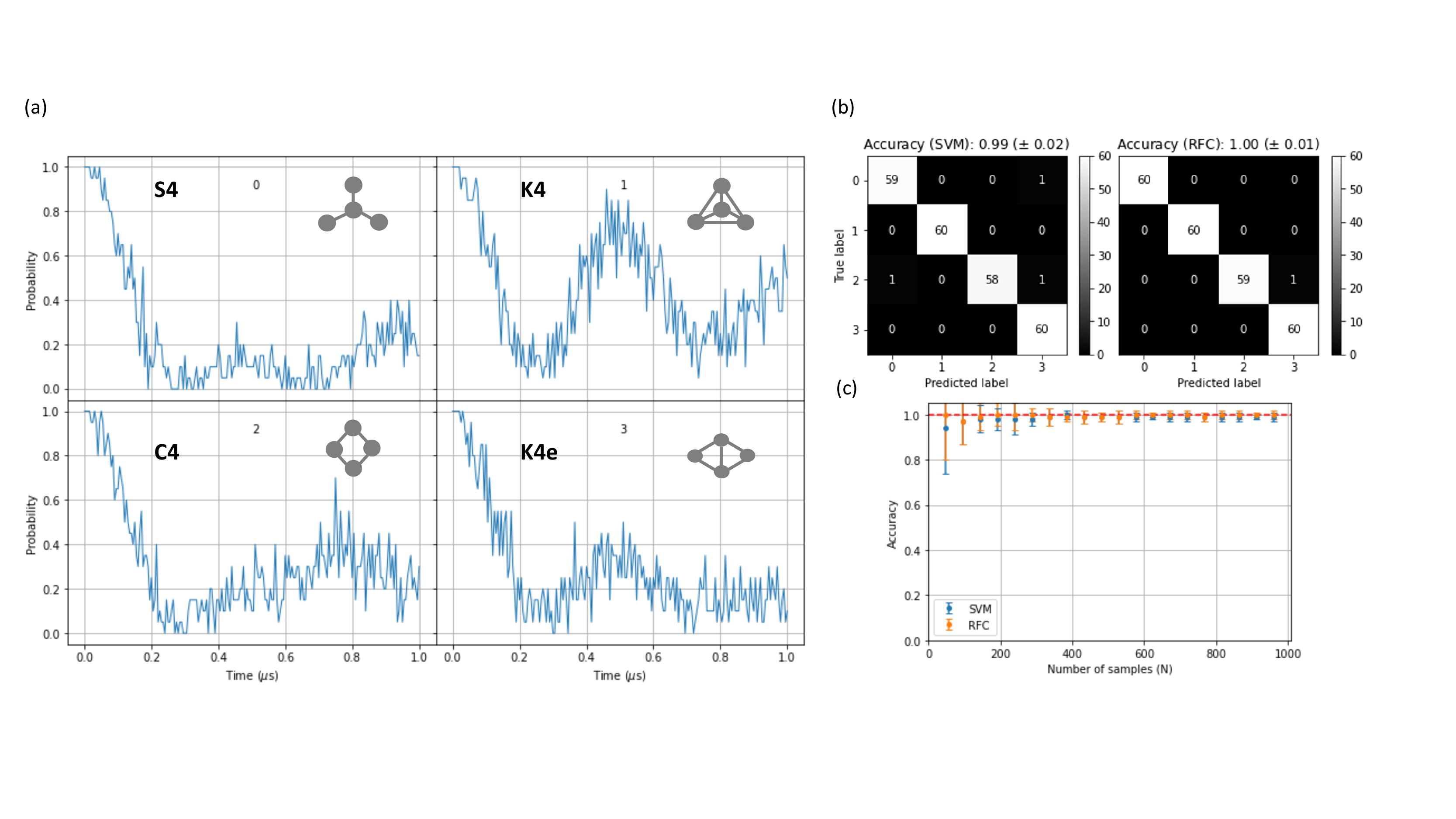}
\caption{Identifying Rydberg ground states of each connected graph for four atoms, $G = $ S4, K4, C4, and K4e. We label each as 0, 1, 2, and 3, respectively. (a) Time evolution of $P_0^G(t) = |\langle \Psi^G(0)|\Psi(t)\rangle|^2= |\langle 0000|\Psi(t)\rangle|^2$ [Eq. (\ref{eq:prob_excited})], and (b) confusion matrix by SVM and RFC methods. (c) Accuracy as a total number of samples $N$. The results show both SVM and RFC accurately predict four graphs due to the distinctive data shapes.}
\label{Fig13}
\end{figure*}


\subsection{Classification by various connected graphs of four ground state atoms: S4, K4, C4, and K4e}

We consider four-atom arrays with various graphs as discussed in Ref. \cite{Kim2020} for supervised training and testing of $N$ sample data. Figure \ref{Fig13} shows the ML multiclass classification of four geometric configurations (connected graph denoted as $G$) in its ground state: S4, K4, C4, and K4e, labeled as 0, 1, 2, and 3, respectively. The time evolution of probabilities is plotted in Fig. \ref{Fig13} (a). To check the performance of the ML  classification, we evaluate the confusion matrix in Fig. \ref{Fig13} (b), indicating the high saturated accuracy of 99\% (2\% error) by SVM and 100\% (1\% error) by RFC. Figure \ref{Fig13} (c) shows that the accuracy reaches its saturation value even before $N~=$ 200. The multiclass classification of four graphs has high accuracy for both models due to the distinctive data patterns in Fig. \ref{Fig13} (a).

So far, we discussed the classification of the number of atoms or shape of a fixed number of atoms in the ground state. For this case, the Rydberg-ID has a one-to-one correspondence with its own interaction Hamiltonian, $U(r_{jk})$, in Eq. (\ref{eq:Hamiltonian}). Next, we extend such multiclass classification to identify possible Rydberg excitation base states. For such cases, the Rydberg-ID does not represent the interaction Hamiltonian unless we form a Rydberg fraction from each base state. 

\subsection{Classification by Rydberg excitation of a fixed configuration.}

\begin{figure*}[tbp!]
  \centering
\includegraphics[width=1.0\textwidth]{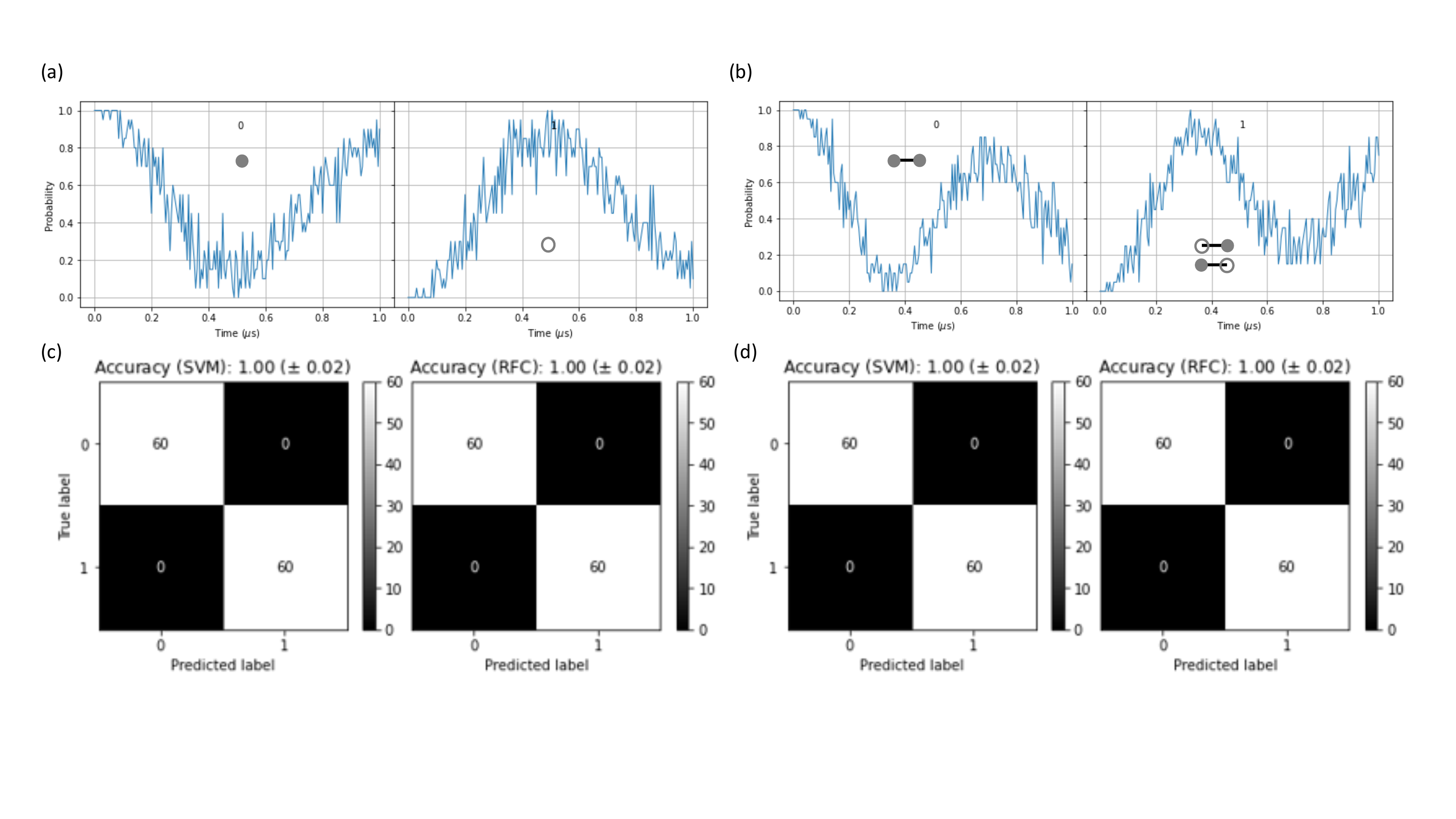}
\caption{One or two atom Rydberg excitation. Both cases consist of two possible base states: $W_0$ (ID = 0) and $W_1$ (ID = 1). Time evolution of probabilities for (a) either $W_0=|0\rangle$ or $W_1=|1\rangle$ in a single qubit atom (S1), and (b) either $W_0=|00\rangle$ or $W_1=(|10\rangle+|01\rangle)/\sqrt{2}$ in two (binary) atoms (B2). Filled grey circles denote atoms in the ground state, and empty circles represent Rydberg excitation of an atom. Confusion matrix of SVM and RFC for (c) S1 and (b) B2 without any confusing decision (no off-diagonal values) corresponding to its highest accuracy.}
\label{Fig2}
\end{figure*}

\begin{figure*}[tbp!]
\centering
  \includegraphics[width=1.05\textwidth]{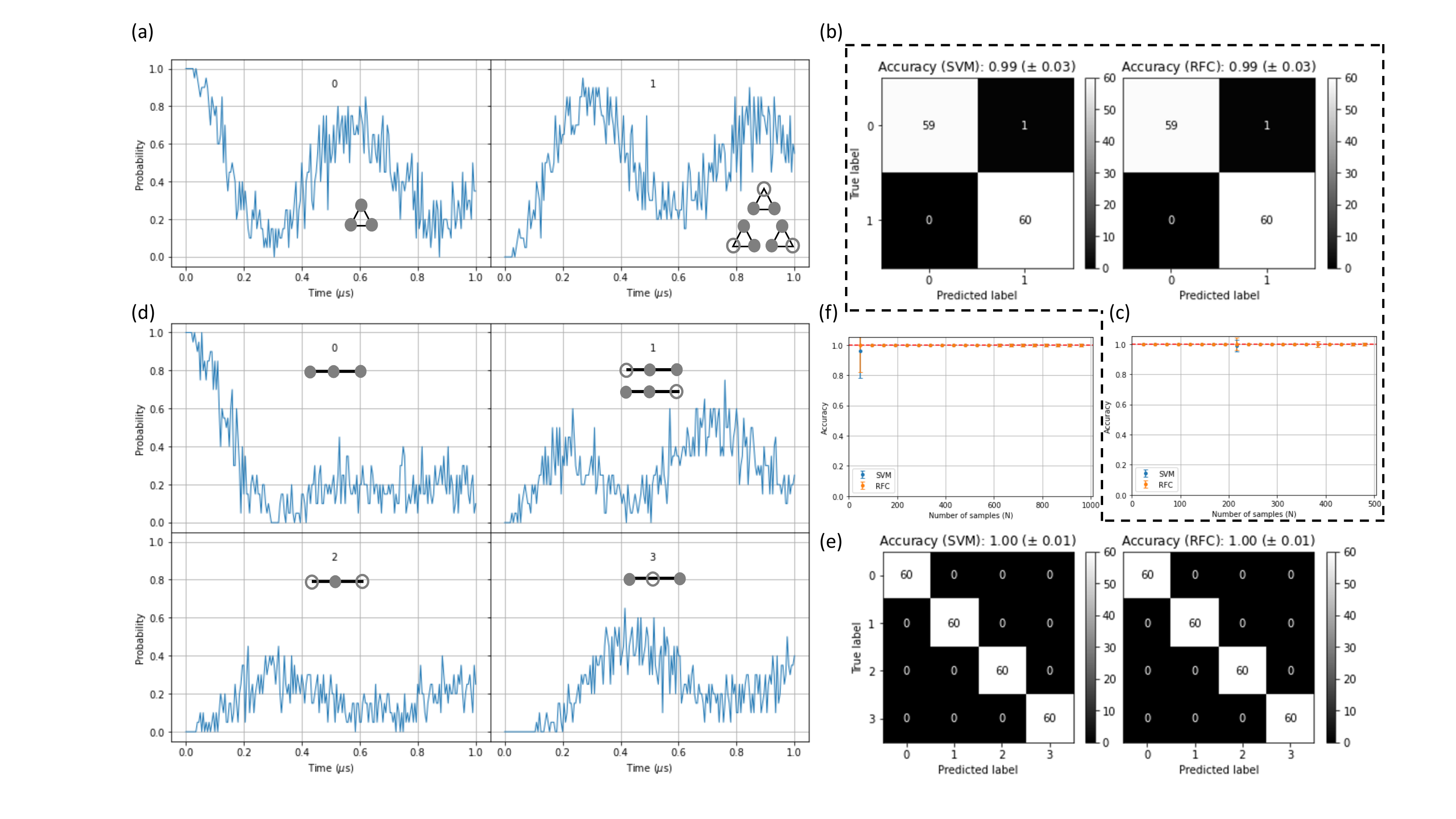}

\caption{Identifying Rydberg states of three atoms in (a-c) triangle shape (T3) resulting in two bases as we label ID = 0 and 1: ground state, 0: $W_0=|000\rangle$, and excited state, 1: $W_1=(|100\rangle+|010\rangle+|001\rangle)/\sqrt{3}$, and (d-f) linear chains consisting of four bases. We label each as ID = 0: $W_0=|000\rangle$, 1: $W_{1a}=(|100\rangle+|001\rangle)/\sqrt{2}$, 2: $W_{1b}=|010\rangle$, and 3: $W_2=|101\rangle$, respectively. $P_{N_{Ryd}}(t) = |\langle \Psi_{N_{Ryd}}|\Psi(t)\rangle|^2$ denotes time evolution of $N_{Ryd}$ number excitation for (a) T3, and (d) linear chains. Confusion matrix from each ML method, \textbf{SVM} and RFC, for classifying all possible bases of $N_{Ryd} = 0$ and $N_{Ryd} = 1$ in (b) T3 and (e) linear chains. Accuracy as a function of sample number $N$ is plotted in (c) and (f), respectively.}
\label{Fig3}
\end{figure*}

\begin{figure*}[tbp!]
  \centering
\includegraphics[width=1.0\linewidth]{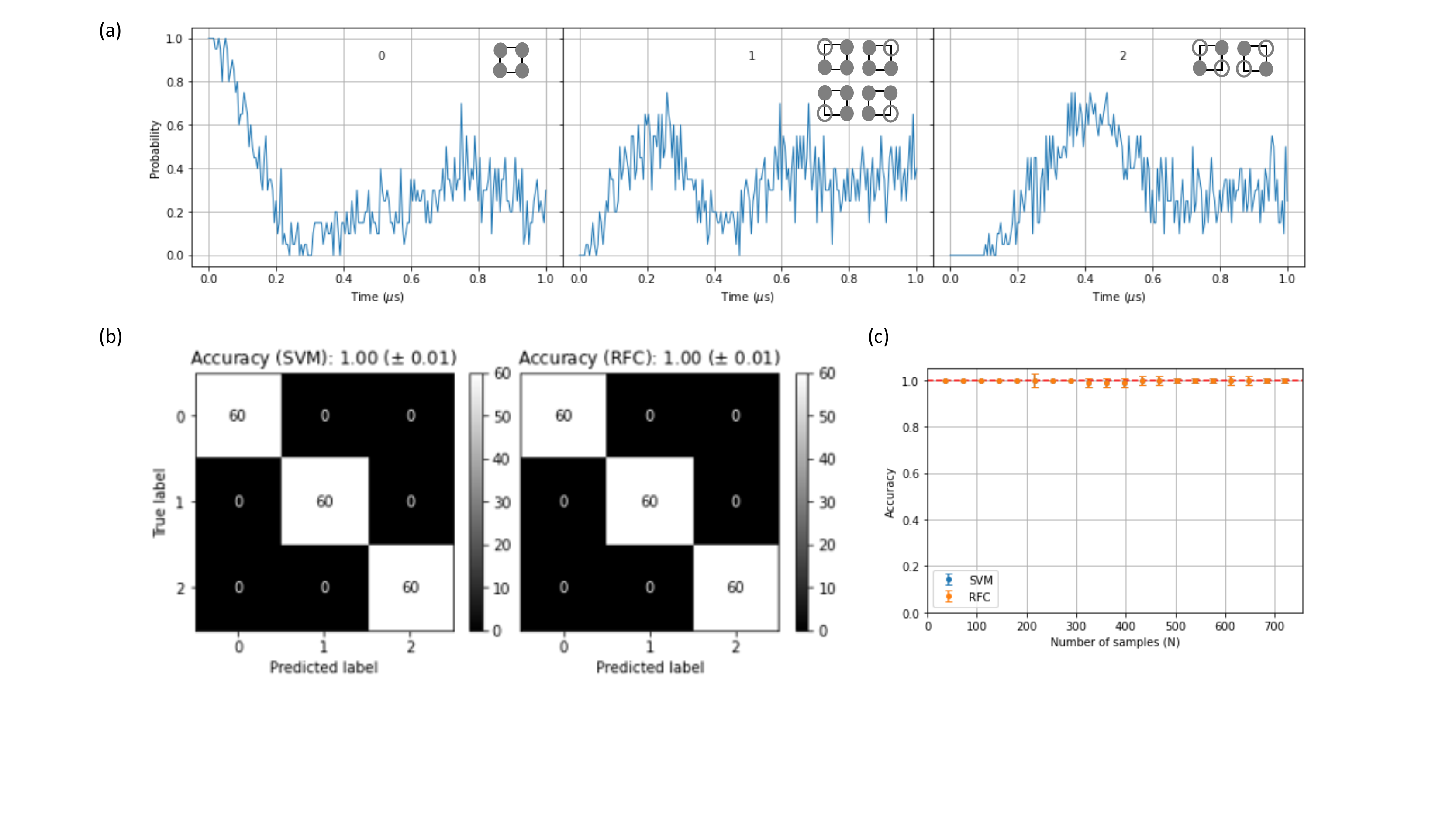}\\
    \includegraphics[width=1.0\textwidth]{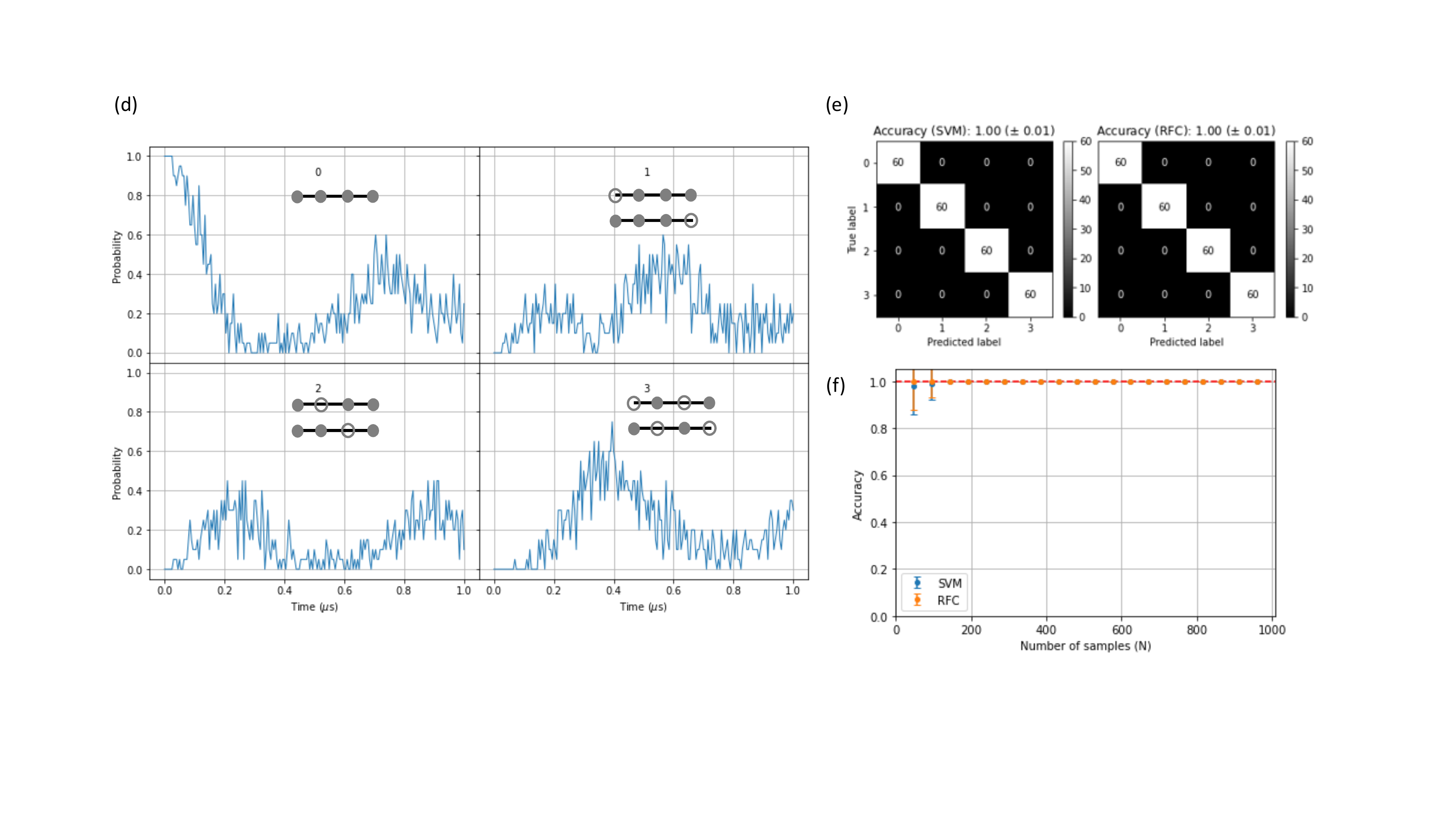}
\caption{Identifying four-atom Rydberg bases of (a-c) square shape (C4) and of (d-f) linear chains. Time evolution $P_{N_{Ryd}}(t) = |\langle \Psi_{N_{Ryd}}|\Psi(t)\rangle|^2$ of the $N_{Ryd}$ excitation for (a) C4. We label each as N = 0: $W_0=|0000\rangle$, 1: $W_1=(|1000\rangle+|0100\rangle+|0010\rangle+|0001\rangle)/\sqrt{4}$, 2: $W_2=(|1010\rangle+|0101\rangle)/\sqrt{2}$. Classification of each shows high accuracy of 100 \% from (b) confusion matrix and (c) accuracy. For linear chains of four atoms, we plot (d) time evolution $P_{N_{Ryd}}(t)$ labeled as 0: $W_0=|0000\rangle$, 1: $W_{1a}=(|1000\rangle+|0001\rangle)/\sqrt{2}$, 2: $W_{1b}=(|0100\rangle+|0010\rangle)/\sqrt{2}$. 3: $W_2=(|1010\rangle+|0101\rangle)/\sqrt{2}$, (e) confusion matrix, and (f) accuracy.}
\label{Fig4}
\end{figure*}

Now, we classify all possible Rydberg base states for each graph. Figure \ref{Fig2} shows Rabi-oscillation of either a single atom [S1, Fig. \ref{Fig2} (a)] or collective Rabi-oscillation of two (binary) atoms [B2, Fig. \ref{Fig2} (b)] with high classification accuracy, which does not require ML classification because of the obvious distinction between data shapes. This trend even lasts for three-atom arrays [Fig. \ref{Fig3}], and four-atom cases [Fig. \ref{Fig4}]. The ML-based classification results in 100\% accuracy even with tens of samples for some cases. For such high accuracy, the data pattern distinctively differs from each other. Except for the single-atom qubit case, multi-qubit atoms are located within the Rydberg blockade range $d=8 ~\mu m$ of the nearest neighbor. 

\begin{figure*}[tbp!]
\centering
\includegraphics[width=0.9\linewidth]{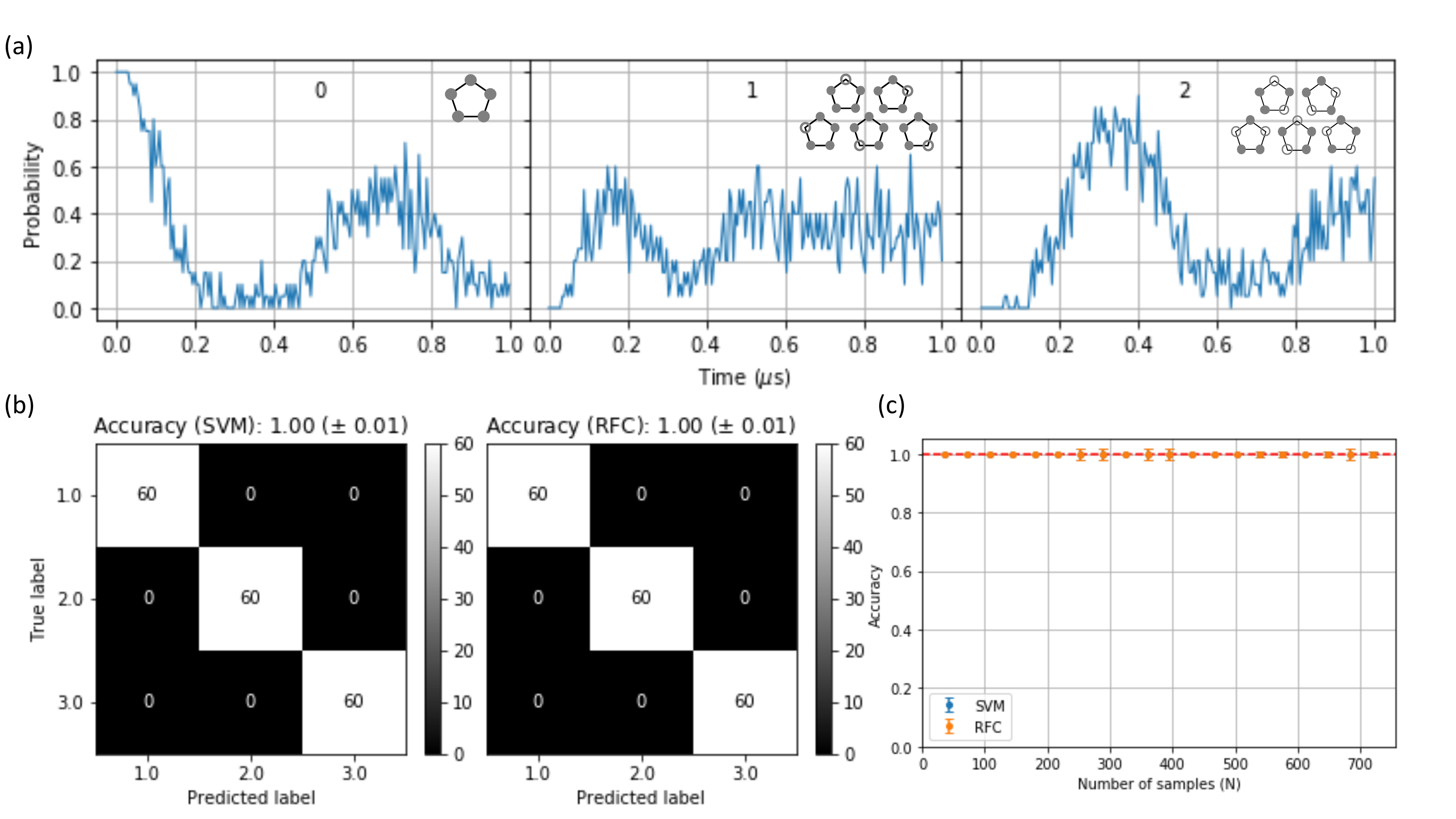}\\
   \includegraphics[width=1.0\textwidth]{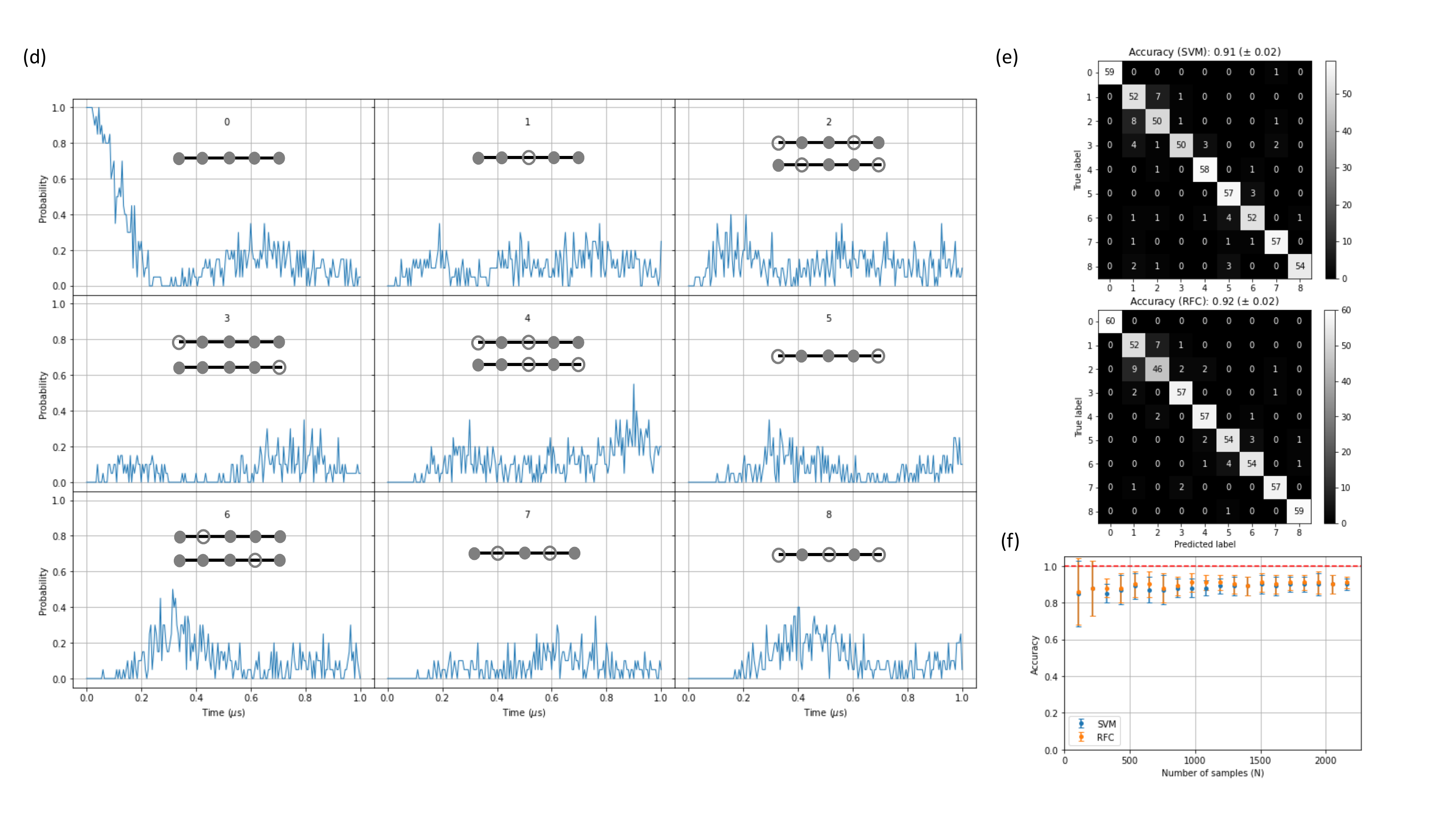}
\caption{Identifying Rydberg ground states of five atoms in (a-c) pentagon shape (P5), and (d-f) linear chains. We label each as 0, 1, 2, ... 8, respectively. (a) Time evolution of $P_{N_{Ryd}}(t) = |\langle \Psi_{N_{Ryd}}|\Psi(t)\rangle|^2$, where the $|\Psi_{N_{Ryd}}\rangle$ denotes a base state of $N_{Ryd}$ number of excitation. We label each as 0: $W_0=|00000\rangle$, 1: $W_1=(|10000\rangle+|01000\rangle+|00100\rangle+|00010\rangle+|00001\rangle)/\sqrt{5}$, 2: $W_2=(|10100\rangle+|01010\rangle+|00101\rangle+|10010\rangle+|01001\rangle)/\sqrt{5}$. (b) Confusion matrix of each SVM and RFC consisting of $N_{Ryd} = 0$, $1$, $2$, and $3$. (c) Accuracy vs. sample number $N$. For linear chains of five atoms, (d) Time evolution $P_{N_{Ryd}}(t)$. We label each as 0: $W_0=|00000\rangle$, 1: $W_{1a}=(|10000\rangle+|00001\rangle)/\sqrt{2}$, 2: $W_{1b}=(|01000\rangle+|00010\rangle/\sqrt{2}$, 3: $W_{1c}=|00100\rangle$, 4: $W_{2a}=(|10100\rangle+|00101\rangle)/\sqrt{2}$, 5: $W_{2b}=|01010\rangle$, 6: $W_{2c}=(|10010\rangle+|01001\rangle)/\sqrt{2}$, 7: $W_{2d}=|10001\rangle$, 8: $W_3=|10101\rangle$,  (e) Confusion matrix. There is a significant difference in the diagonal value (true positive) of 3 for the two methods. (f) Accuracy. The saturated accuracy values reach 91\% $\sim$ 92\%.}
\label{Fig5}
\end{figure*}

\begin{figure*}[tbp!]
  \centering
\includegraphics[width=0.9\linewidth]{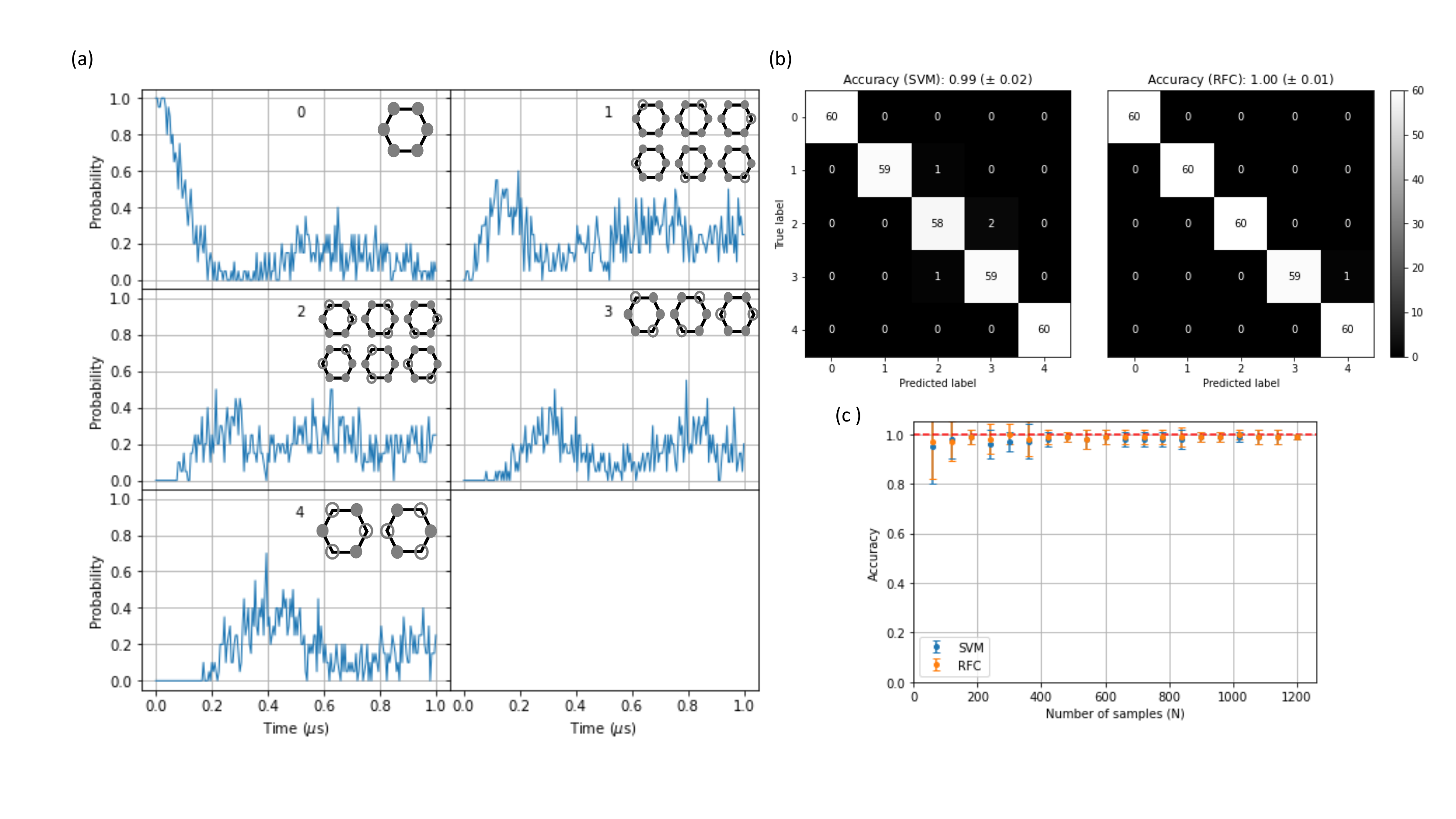}\\
    \includegraphics[width=0.99\textwidth]{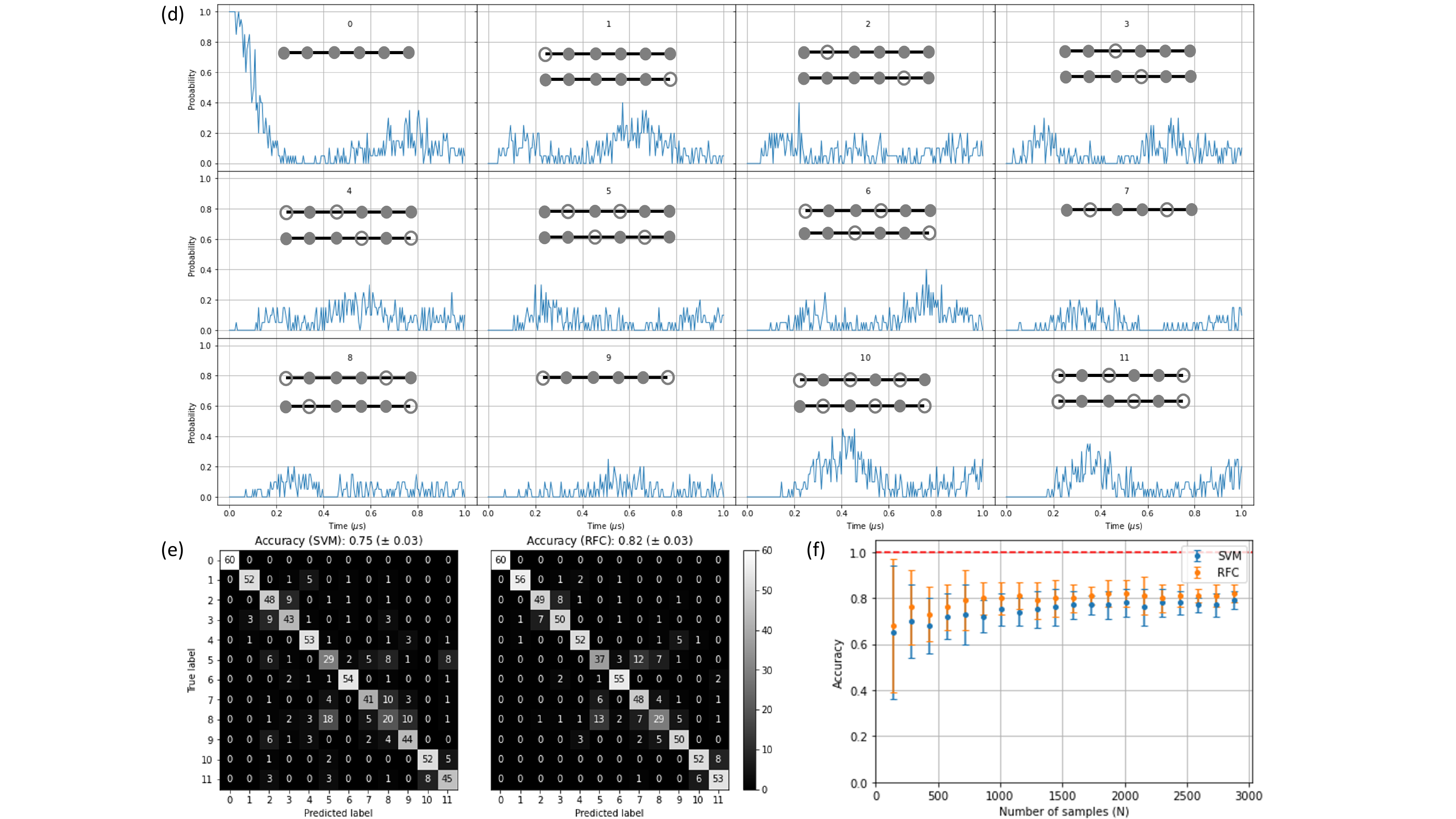}
\caption{Identifying Rydberg excitation configuration of six atoms in (a-c) hexagonal shape (H6), and (d-f) linear chains. We label each as 0, 1, 2, ... 11, respectively. (a) Time evolution of $P_{N_{Ryd}}(t) = |\langle \Psi_{N_{Ryd}}|\Psi(t)\rangle|^2$, where the $|\Psi_{N_{Ryd}}\rangle$ denotes base states of $N_{Ryd}$ number of excitation. We label each as 0: $W_0=|000000\rangle$, 1: $W_1=(|100000\rangle+|010000\rangle+|001000\rangle+|000100\rangle+|000010\rangle+|000001\rangle)/\sqrt{6}$, 2: $W_{2a}=(|101000\rangle+|010100\rangle+|001010\rangle+|000101\rangle+|100010\rangle+|010001\rangle)/\sqrt{6}$, 3: $W_{2b}=(|100100\rangle+|010010\rangle+|001001\rangle)/\sqrt{3}$, 4: $W_3=(|101010\rangle+|010101\rangle)\sqrt{2}$. (b) confusion matrix of SVM and RFC consisting of $N_{Ryd} = 0$, $1$, $2$, and $3$. (c) Accuracy as a number of samples $N$. (d) Time evolution $P_{N_{Ryd}}(t)$ for linear chains.  We label each as 0: $W_0=|000000\rangle$, 1: $W_{1a}=(|100000\rangle+|000001\rangle)/\sqrt{2}$, 2: $W_{1b}=(|010000\rangle+|000010\rangle/\sqrt{2}$, 3: $W_{1c}=(|001000\rangle+|000100\rangle/\sqrt{2}$, 4: $W_{2a}=(|101000\rangle+|000101\rangle/\sqrt{2}$, 5: $W_{2b}=(|010100\rangle+|001010\rangle)/\sqrt{2}$, 6:
$W_{2c}=(|100100\rangle+|001001\rangle)/\sqrt{2}$, 7:
$W_{2d}=|010010\rangle$, 8: $W_{2e}=(|100010\rangle+|010001\rangle)/\sqrt{2}$, 9: $W_{2f}=|100001\rangle$, 10: $W_{3a}=(|101010\rangle+|010101\rangle)/\sqrt{2}$, 11: $W_{3b}=(|101001\rangle+|100101\rangle)/\sqrt{2}$. (e) Confusion matrix indicates mixed off-diagonal elements  between ``2'' ($W_{1b}$) and ``3'' ($W_{1c}$), ``5'' ($W_{2b}$) and ``8'' ($W_{2e}$), ``7'' ($W_{2d}$) and ``8'' ($W_{2e}$), ``8'' ($W_{2e}$)and ``9'' ($W_{2f}$), and ``10'' ($W_{3a}$) and ``11'' ($W_{3b}$). (f) Consequently, the accuracy level is around 80 \%.}
\label{Fig6}
\end{figure*}

For more than three-atom arrays, we locate atoms in two different ways; closed configurations and linear chains, as shown in Fig. \ref{Fig3} - Fig. \ref{Fig6}. Specifically, we classify symmetric triangle (T3) [Fig. \ref{Fig3} (a-c)] and equidistant linear chains [Fig. \ref{Fig3} (d-f)]. Two possible bases of T3 are the ground state $|000\rangle$ and excited state $(|100\rangle+|010\rangle+|001\rangle)/\sqrt{3}$, respectively labeled as 0 and 1. Such symmetric configuration for closed shapes are applied to higher number of atoms discussed later: square (C4) [Fig. \ref{Fig4} (a-c)], pentagon (P5) [Fig. \ref{Fig5} (a-c)], hexagon (H6) [Fig. \ref{Fig6} (a-c)]. 

The three-atom linear chain results in four possible Rydberg base states labeled as ID = 0, 1, 2, and 3. The number of of atoms in the linear chain is increased up to six later on [Fig. \ref{Fig4}-\ref{Fig6} (d-f)]. $P_{N_{Ryd}}(t) = |\langle \Psi_{N_{Ryd}}|\Psi(t)\rangle|^2$ denotes time evolution of $N_{Ryd}$ number excitation. Fig. \ref{Fig3} (b) and (e) shows the confusion matrix. We plot the accuracy as a function of sample number N = (0, $N_{ID} \times $300) in Fig. \ref{Fig3} (c) and (f) to estimate the sample size needed to reach the saturation accuracy. The probability plots of T3 and three-atom linear chains are well distinguished from each other, which results in high accuracy for both models.

Figure \ref{Fig4} shows the four-atoms case similar to the three atom case. Up to four atoms, the accuracy reaches 100\% with fast saturation even with few training samples for both closed square shape (C4) and linear chains. As we increase the number of atoms, however, the behavior of accuracy changes in some cases. It is because the corresponding time evolution data $P_{N_{Ryd}}(t)$ becomes more indistinguishable, and the patterns get complicated, especially for the linear chains. 

The closed shape of the five-atom configuration (pentagon, P5) has three base states corresponding to the time evolution data in Fig. \ref{Fig5} (a). Due to the distinct shapes, the accuracy is 100\% (1\% error) with only true positives (TP) or diagonal elements for both SVM and RFC in Fig. \ref{Fig5} (b). Consequently, Fig. \ref{Fig5} (c) shows constant 100\% accuracy regardless of the training sample size $N$. 

Figure \ref{Fig5} (d) shows the probability of five-atom linear chains with each graphic configuration, respectively. The confusion matrix in Fig. \ref{Fig5} (e) has non-zero off-diagonal values indicating the false-positives (FP) and false-negatives (FN) of our models.
Such errors arise from similar probability profiles, for example, between ID ``1'' ($W_{1a}=(|10000\rangle+|00001\rangle)/\sqrt{2}$) and ``2'' ($W_{1b}=(|01000\rangle+|00010\rangle)/\sqrt{2}$) in Fig. \ref{Fig5} (d). The notable difference between SVM and RFC is the diagonal value of ``3''. The single model method (SVM) shows 50, and the ensemble method (RFC) has 57 true positives (TP). The origin of the SVM model error seems to be the mixture of ``1'' and ``3'' compared to RFC. It would be one example to see how those two multiclass classifiers work from the data shape in Fig. \ref{Fig5} (d). Nonetheless, the accuracy cannot reach 100 \%. The saturated accuracy of RFC reaches up to 92 \% (2\% error) instead [Fig. \ref{Fig5} (e)-(f)]. 

A hexagonal array of six atoms (H6) is the most representative closed system for various multi-partite entangled systems as shown in Fig. \ref{Fig6} (a). Because of the clear distinction among data, the confusion matrix's main diagonal values in Fig. \ref{Fig6} (b) guarantee both models' classification accuracy up to 100\%. However, the linear counterpart in Fig. \ref{Fig6} (d) shows otherwise with saturation accuracy around 80\% [Fig. \ref{Fig6} (e)-(f)]. Interestingly, the confusion matrix in Fig. \ref{Fig6} (e) shows acceptable main diagonal values (TPs) except for significant mixed off-diagonal elements between ``2'' ($W_{1b}$) and ``3'' ($W_{1c}$), ``5'' ($W_{2b}$) and ``8'' ($W_{2e}$), ``7'' ($W_{2d}$) and ``8'' ($W_{2e}$), ``8'' ($W_{2e}$) and ``9'' ($W_{2f}$), and ``10'' ($W_{3a}$) and ``11'' ($W_{3b}$). For example, such off-diagonal properties (FNs-FPs) appear among certain doubly excited Rydberg states (ID = 5, 7, 8, and 9) [Fig. \ref{Fig6} (e)]. These are also associated with the indistinguishable data patterns in Fig. \ref{Fig6} (d). 

\section{Discussion}
In the previous sections, we use two supervised classification algorithms, SVM and RFC, implemented in scikit-learn \cite{HOML} to identify Rydberg interaction  Hamiltonians based on the graph of the $^{87}$Rb-single atom array. In this section, we further test accuracy beyond the categories discussed.

\subsection{Accuracy dependence on nearest neighbor interaction and combined closed-linear chains.}

\begin{figure*}[tbp!]
  \centering
\includegraphics[width=1.0\linewidth]{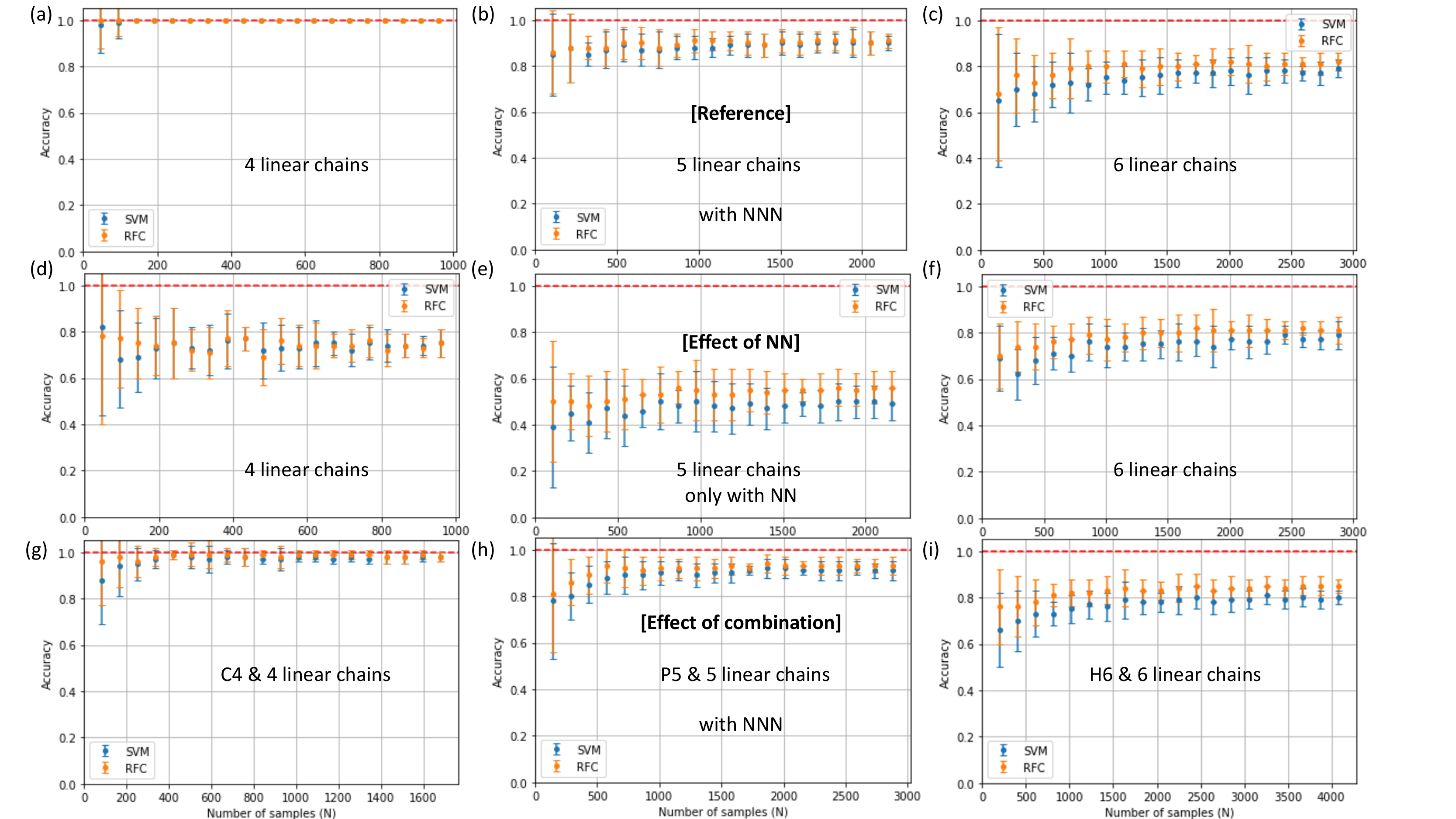}
\caption{Effect of interaction or mixture on accuracy as a function of data samples $N$ for linear chains. The reference linear chains from Fig.\ref{Fig4} - Fig.\ref{Fig6} are shown for (a) four-atom linear chains, (b) five-atom linear chains, and (c) six-atom linear chains with the default next nearest neighbor interactions for comparison with other effects. Accuracy drops significantly when we only consider nearest neighbor (NN) interactions for (d) four-atom linear chains, (e) five-atom linear chains. The accuracy of (f) six-atom linear chains is relatively unaffected. It shows that the next nearest neighbor consideration (a-c) demonstrates high accuracy compared to only the nearest interaction cases (d-f). We further compared (a-c) to the combined shapes, including both closed and linear chains in (g-i), which follows the accuracy of the reference linear chains in (a-c). Note that the above two factors do not really affect the six-atom cases that appeared in (c), (g), and (i).}
\label{Fig7}
\end{figure*}

To simulate experimental data reasonably, we include the next nearest neighbor (NNN) interaction in addition to the nearest neighboring (NN) atom interaction for all $^{87}$Rb atom arrays discussed above. We note that the NNN approximation also improves the ML classifiers' accuracy. For example, we compare the classification accuracy of the NNN and NN only interaction cases as well as combined sets in Fig. \ref{Fig7}.

From Fig. \ref{Fig4} - Fig. \ref{Fig6}, we have seen the high accuracy for the closed shapes compared to their linear counterparts of four-, five-, and six-atom linear chains. Figure \ref{Fig7} (a-c) show the accuracy plots for linear chains as our reference to compare with other effects. Fig. \ref{Fig7} (d-f) shows that the accuracy drops significantly when considering only the nearest neighbor interaction, except for the six-atom linear chain, compared to when the NNN interaction was included in Fig. \ref{Fig7} (a)-(c). Therefore, considering the next nearest neighbor interaction represents the actual experimental data and improves the performance of the ML classifiers.

We further test ML accuracy by combining both closed and linear shapes as in Fig. \ref{Fig7} (g-i). It shows that the mixture causes relatively similar classification accuracy to the linear chain cases in Fig. \ref{Fig7} (a-c). It means that the linear chains' classification characters limit the accuracy of the mixed set of closed and linear chains.

\begin{figure*}[tbp!]
  \centering
\includegraphics[width=0.95\textwidth]{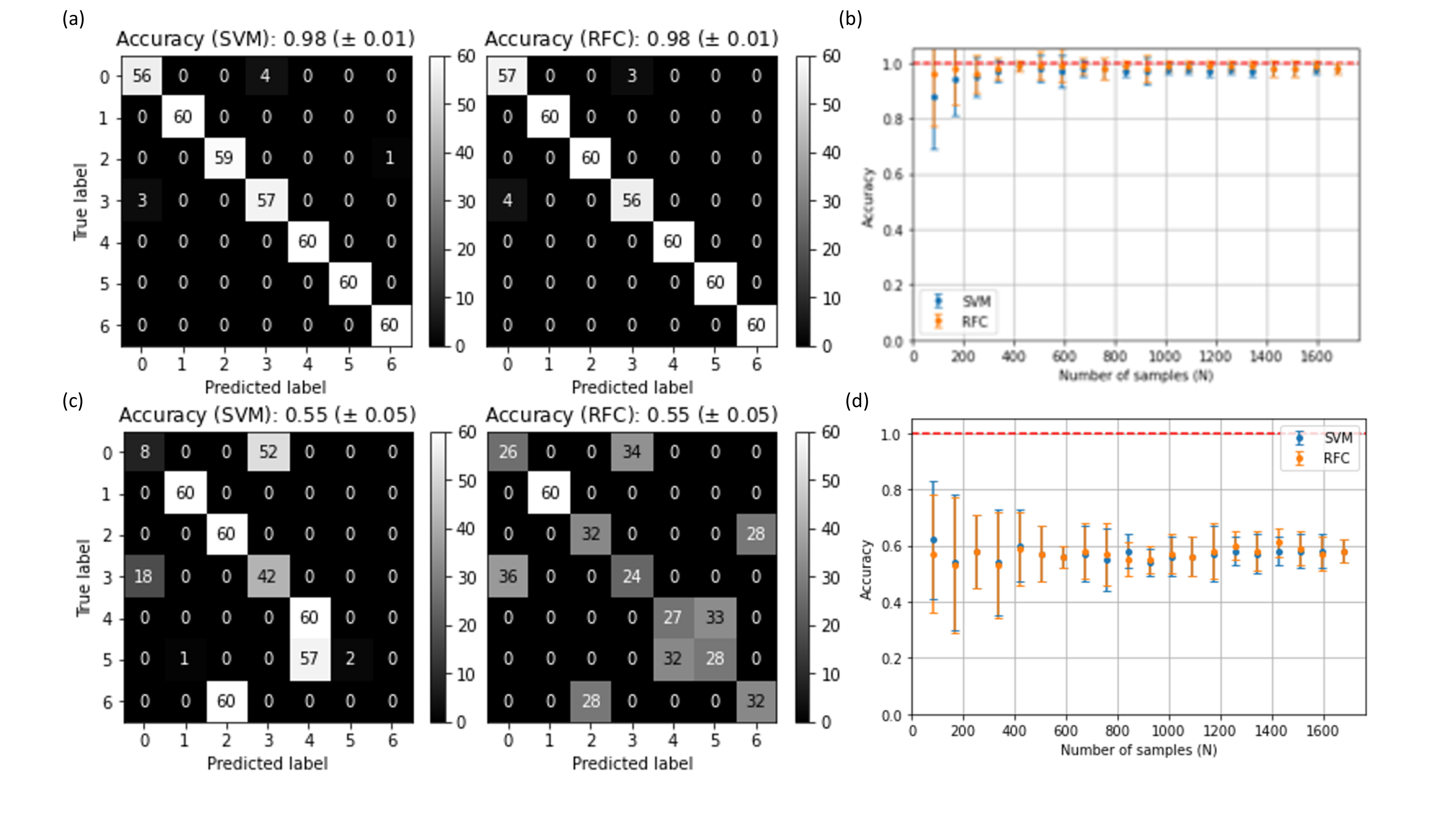}
\caption{
Identification of all seven combined configurations of both C4 and four-atom linear chain. IDs are 0: $W^{C4}_{0}$, 1: $W^{C4}_{1}$, 2: $W^{C4}_{2}$, 3: $W_{0}$, 4: $W_{1a}$, 5: $W_{1b}$, and 6: $W_{2}$.
Confusion matrix for (a) next nearest neighbor (NNN) interaction and (c) nearest neighbor (NN) interaction that shows significant confusion between the ground state of each C4 and linear chains. Accuracy against sample numbers for (b) NNN interaction and (d) NN interaction. 
Consequently, for our default NNN, we see dramatically enhanced accuracy from 55\% to 98\% for both SVM and RFC.}
\label{Fig4-3}
\end{figure*}

\begin{figure*}[tbh!]
\centering
\includegraphics[width=0.9\textwidth]{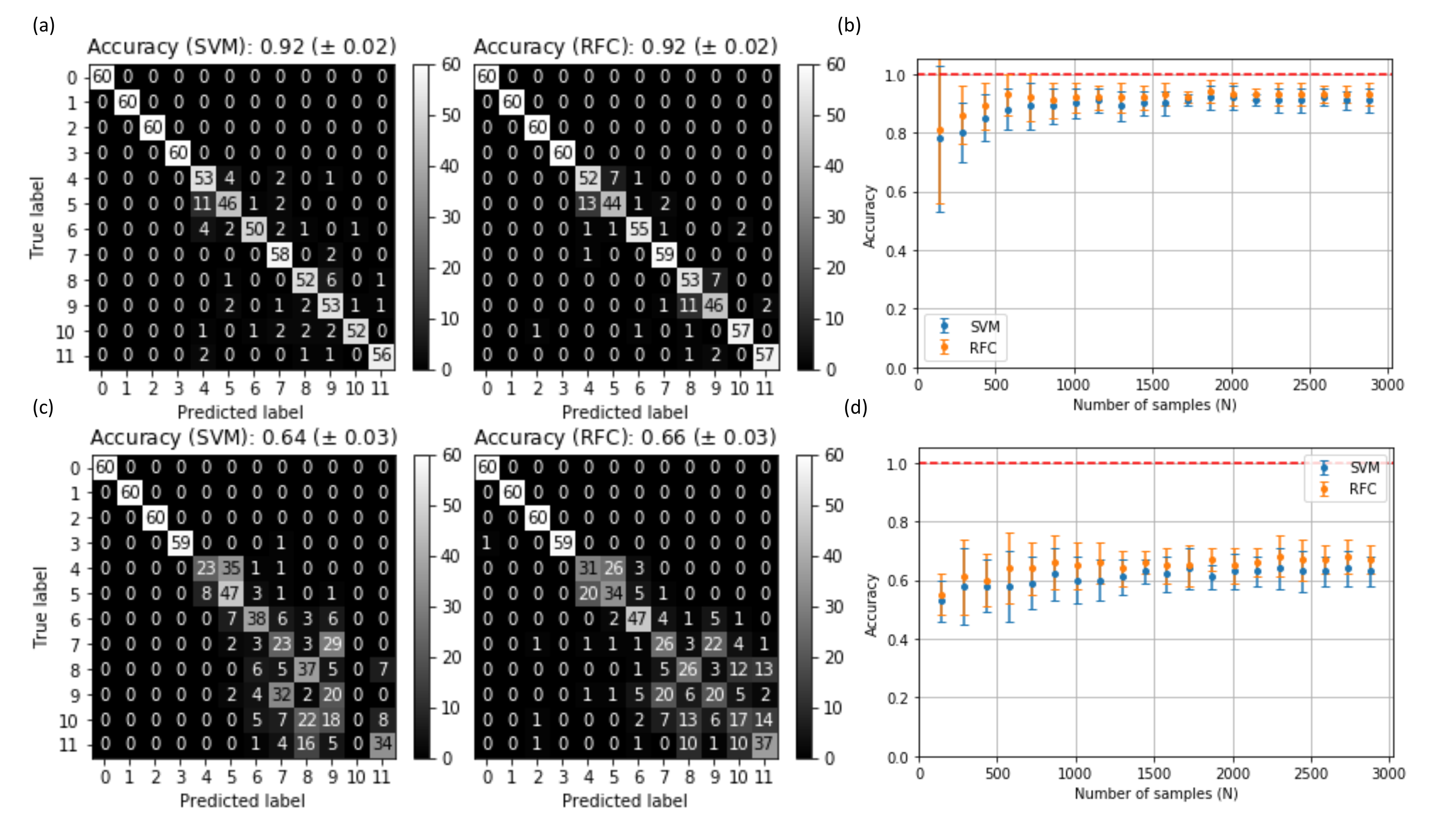}
\caption{Identification of all 12 configurations of both P5 and five-atom linear chain. IDs are 0: $W^{P5}_{0}$, 1: $W^{P5}_{1}$, 2:  $W^{P5}_{2}$, 3: $W_{0}$, 4: $W_{1a}$, 5: $W_{1b}$, 6: $W_{1c}$, 7: $W_{2a}$, 8: $W_{2b}$, 9: $W_{2c}$, 10: $W_{2d}$, and 11: $W_{3}$. Confusion matrix for (a) next nearest neighbor (NNN) interaction and (c) nearest neighbor (NN) interaction. Accuracy for (b) NNN interaction and (d) NN interaction.
We see dramatically enhanced accuracy when the NNN interaction is considered, especially for IDs between 4 and 11.}
\label{Fig5-3}
\end{figure*}

\begin{figure*}[tbh!]
  \centering
\includegraphics[width=0.9\textwidth]{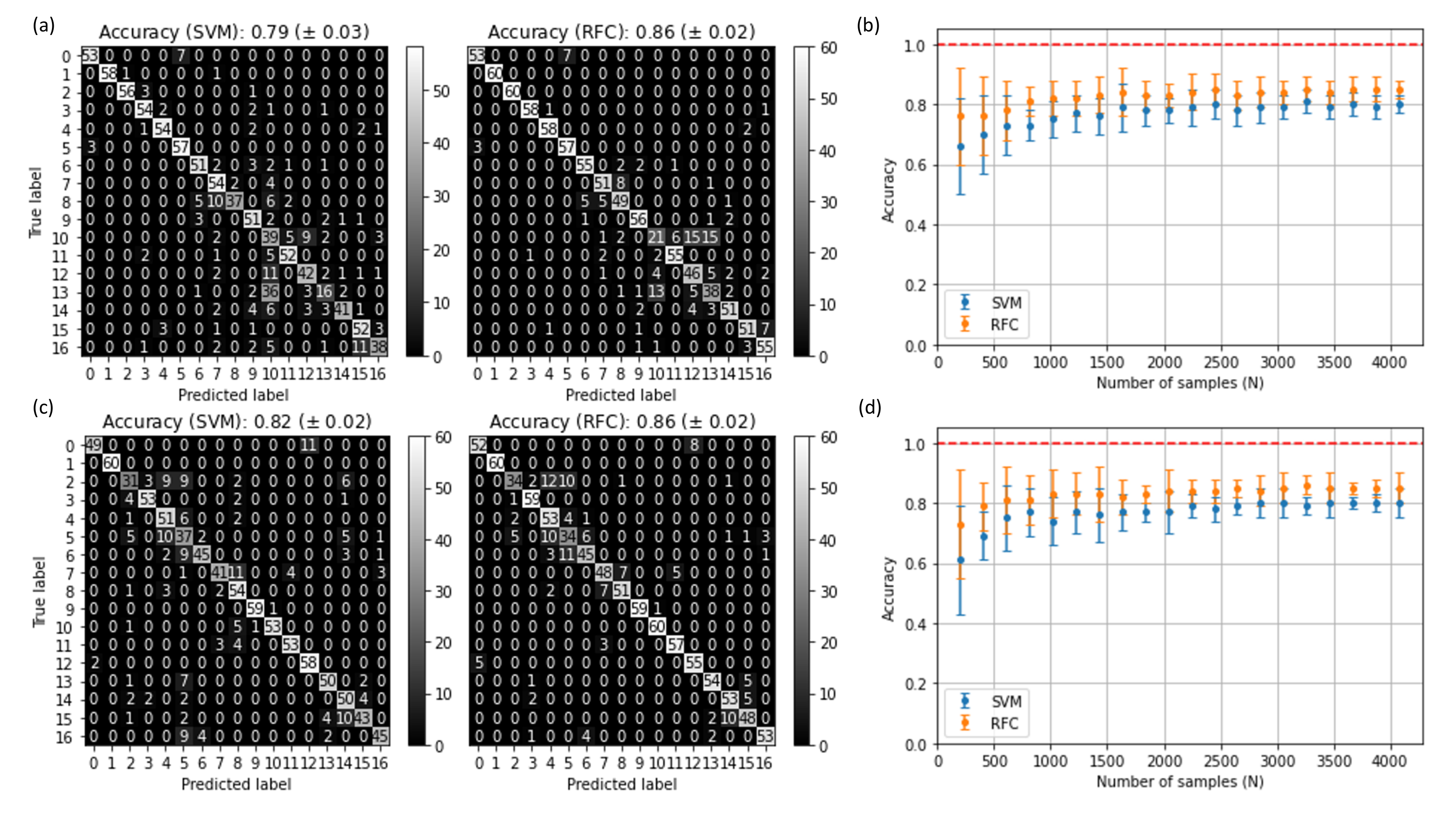}
\caption{Identification of all 17 configurations of both H6 and six-atom linear chain. IDs are 0: $W^{H6}_{0}$, 1: $W^{H6}_{1}$, 2:  $W^{H6}_{2a}$, 3: $W^{H6}_{2b}$, 4: $W^{H6}_{3}$, 5: $W_{0}$, 6: $W_{1a}$, 7: $W_{1b}$, 8: $W_{1c}$, 9: $W_{2a}$, 10: $W_{2b}$, 11: $W_{2c}$, 12: $W_{2d}$, 13: $W_{2e}$, 14: $W_{2f}$, 15: $W_{3a}$, and 16: $W_{3b}$. Confusion matrix for (a) next nearest neighbor (NNN) interaction and (c) nearest neighbor (NN) interaction. Accuracy against sample numbers $N$ for (b) NNN interaction and (d) NN interaction.
Interestingly, in contrast to the four and five atom configurations, adding the NNN interaction doesn't change the accuracy that much within the uncertainty range.
}
\label{Fig6-4}
\end{figure*}

For more details, the combined sets are tested for both the default next nearest neighbor (NNN) interaction and the nearest neighbor (NN) interaction for four, five, and six atoms from Fig.\ref{Fig4-3} to Fig.\ref{Fig6-4}. The default NNN interaction results in the confusion matrix's main diagonal values similar to the case of linear chains discussed in Fig.\ref{Fig4}-Fig.\ref{Fig6}. 

Figure \ref{Fig4-3} (a) shows the confusion matrix of all the seven base states from the combined C4 and four-atom linear chains. We show the accuracy plot in Fig. \ref{Fig4-3} (b), which is the same as Fig. \ref{Fig7} (g). The same base states with only NN interaction are shown in Fig. \ref{Fig4-3} (c) confusion matrix, and (d) the accuracy plot, which shows drastically degraded classification performance from 55\% compared to 98\% for NNN. Figure \ref{Fig4-3} (c), for example, shows erroneous decision between ground states of C4 ($W_0^{C4}$, ID = 0) and four-chains ($W_0$, ID = 3). The next nearest neighbor (NNN) interaction seems to be the essential part of rendering the linear chains' physical character in this case. 

For the 12 cases of the combined P5 and five-chains, the confusion matrix in Fig. \ref{Fig5-3} (a) and the accuracy plot in Fig. \ref{Fig5-3} (b) follow the pattern of the linear chains as discussed in Fig. \ref{Fig7} (b) and (h). However, with only the nearest neighbor (NN) interactions in the linear chains, we see wrong predictions among ID = 4 $\sim$ 11, which is the linear-chain part. When we consider the default NNN, the confusion matrix keeps its diagonal character representing high classification performance.

Figure \ref{Fig6-4} shows how NNN or NN affects the confusion matrix and accuracy for identifying all 17 base states from the combined H6 and its linear counterpart. The performance of both, however, does not differ much. Natural questions that arise are ``Why does the NNN correction term enhance the accuracy dramatically for four and five-atom linear chains?'' and ``Does it indicate intermediate regime at a certain number of linear chains?''. These questions indicate that ML classification could be useful to study the Rydberg Hamiltonian and base states in detail in the future.

\subsection{Noise effect on accuracy}

\noindent For the data in the previous sections, we considered 3\% laser intensity noise fluctuation level and its 1\% standard deviation based on our experience in taking real experimental data. To see how the ML accuracy depends on the noise, we increase the average noise level and its standard deviation to 7\% (2\%) and 10\% (5\%).

\begin{figure*}[tbh!]
  \centering
\includegraphics[width=1.00\linewidth]{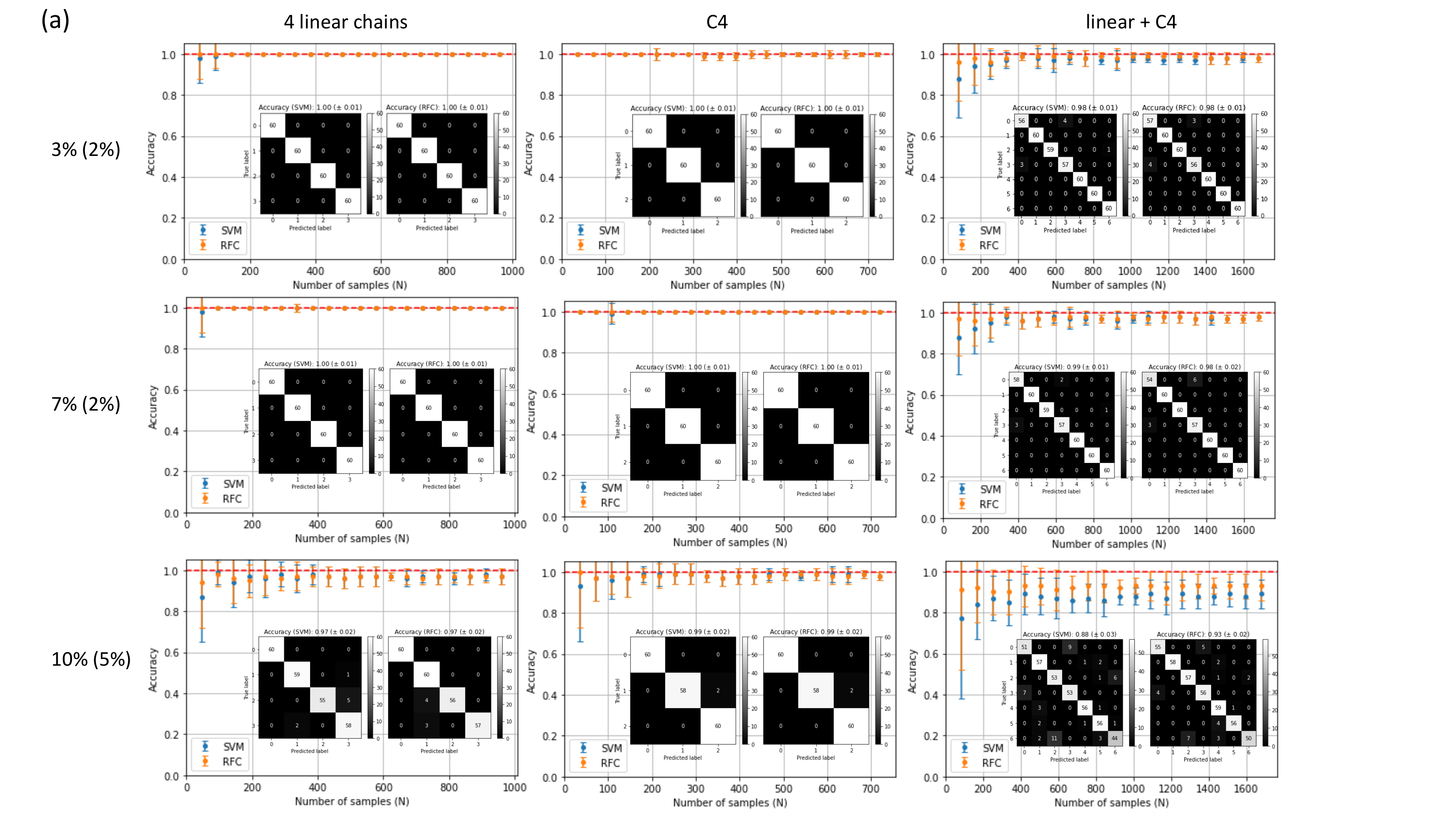}\\ 
\includegraphics[width=0.9\linewidth]{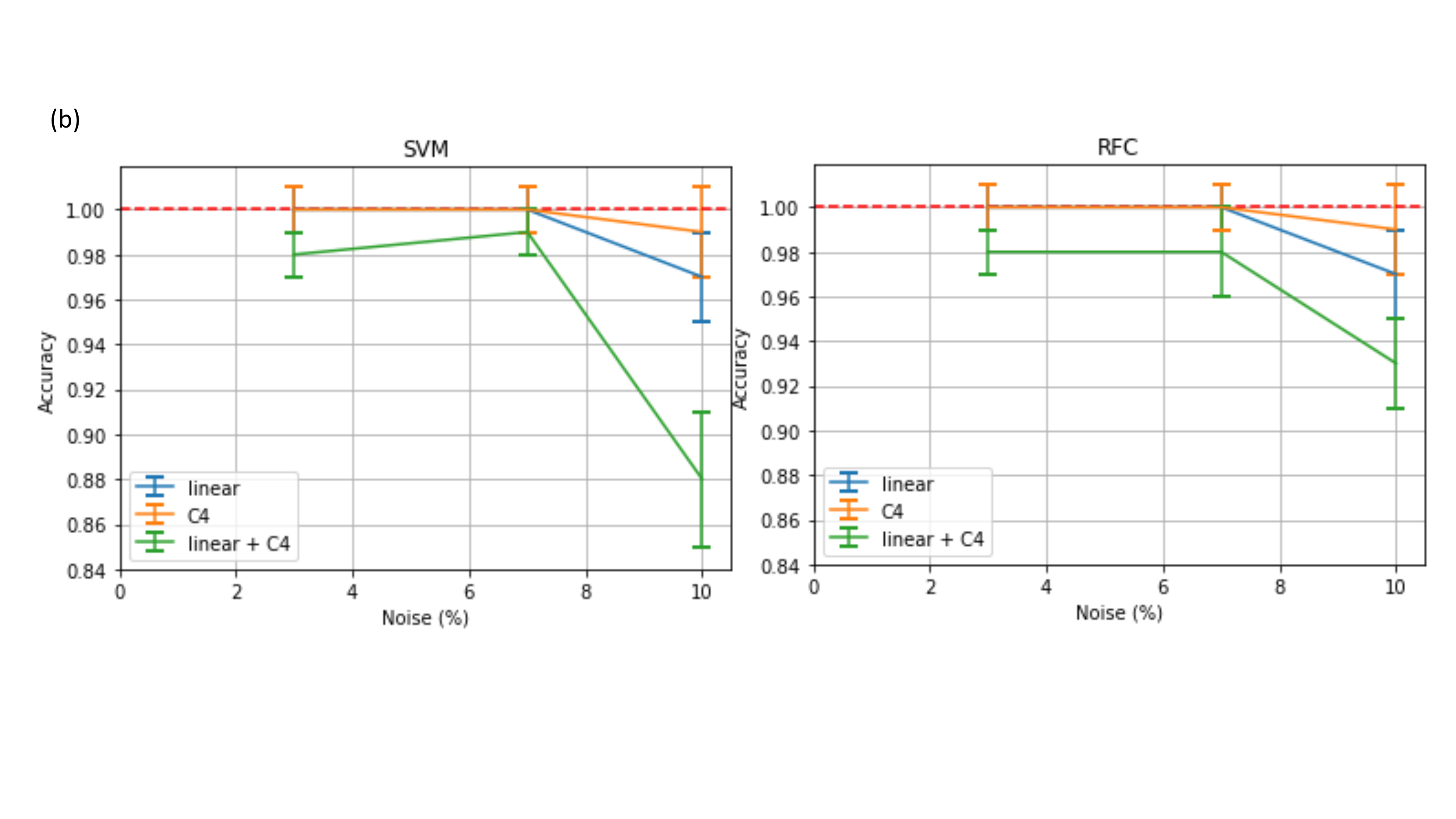}
\caption{Accuracy dependence on laser noise intensity for four atoms. (a) Accuracy as a function of sample size $N$ for various laser noise intensity fluctuations. (b) Accuracy as a function of fluctuation level (standard deviation). The fluctuation of 3\% (1\%) and 7\% (2\%) represent practical experimental situations in the lab. For comparison, we considered the significant fluctuation level of 10\% (5\%), which rarely occurs.}
\label{Fig14}
\end{figure*}

For four atoms, Fig. \ref{Fig14} (a) shows the effect of noise fluctuation on the accuracy. The accuracy does not decrease much for both realistic experimental situations of 3\% (1\%) and 7\% (2\%). However, increasing the mean value to 10\% (5\%) drops the accuracy to 97 $\sim$ 98\% for closed and linear chains, still showing high accuracy except for the combined case. For the mixed case, the accuracy drops dramatically to 88 \% for SVM and 93\% for RFC. Fig. \ref{Fig14} shows accuracy as a function of (a) sample size and (b) the noise (fluctuation). Increasing the noise seems to confuse data shapes further for the mixture of the closed and linear chains.

\begin{figure*}[tbh!] 
    \centering
    \includegraphics[width=1.0\linewidth]{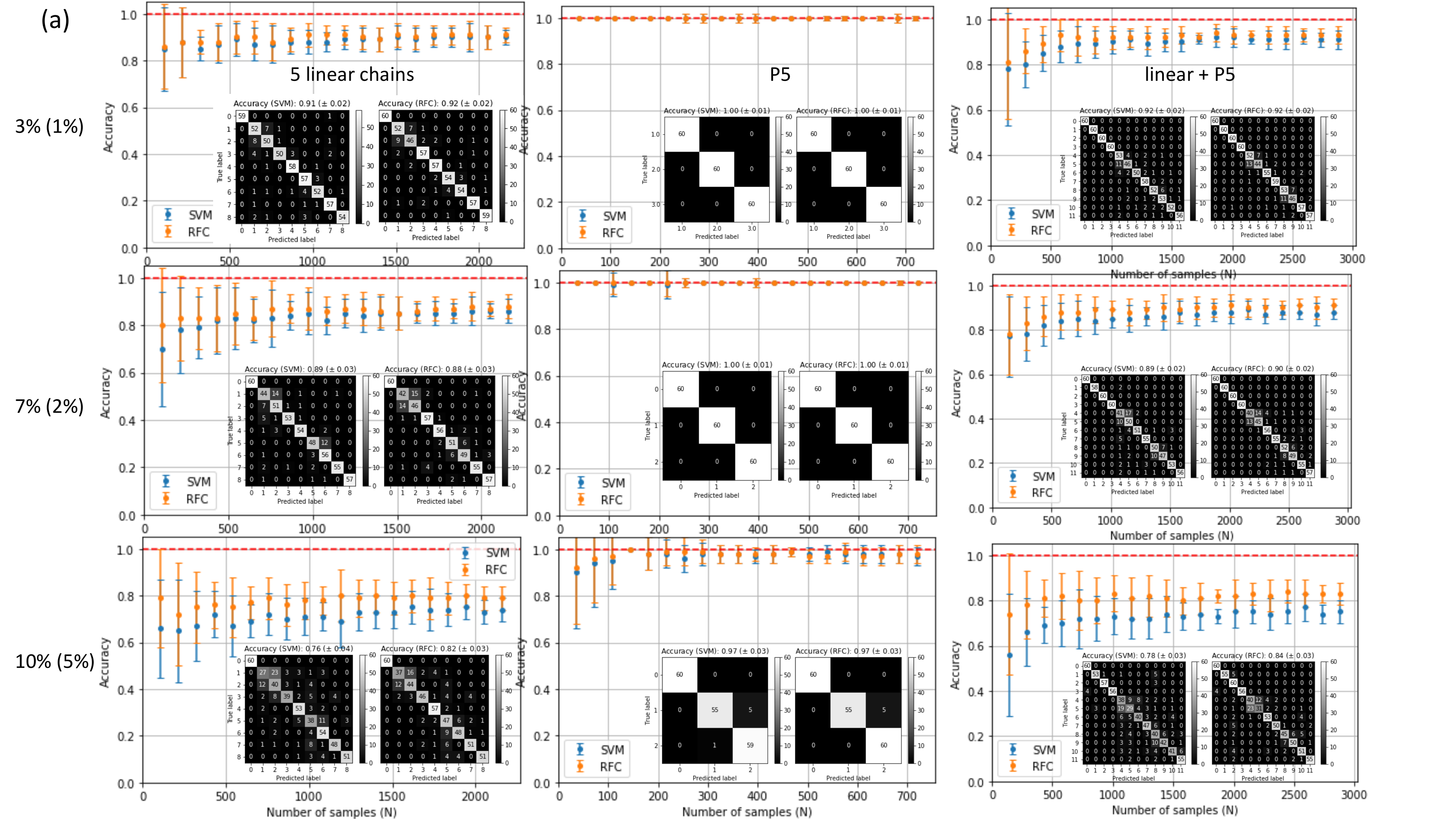}\\
    \includegraphics[width=0.9\linewidth]{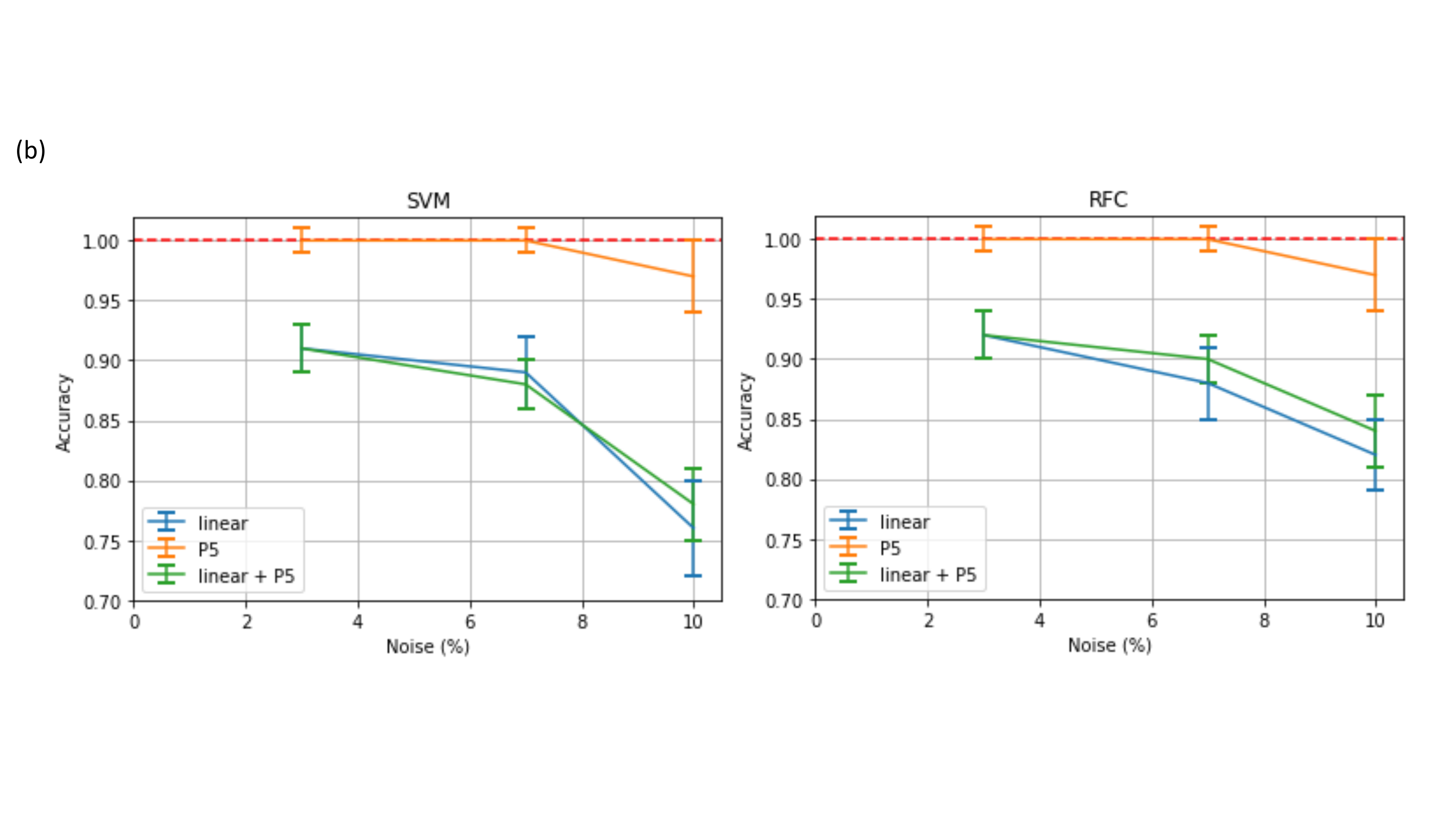}
    \caption{Accuracy dependence on laser noise intensity for five atoms. (a) Accuracy as a function of sample size $N$ for various laser noise intensity fluctuations. (b) Accuracy as a function of fluctuation level (standard deviation). The fluctuation of 3\% (1\%) and 7\% (2\%) represent practical experimental situations in the lab. For comparison, we considered the significant fluctuation level of 10\% (5\%), which rarely occurs.}
    \label{Fig15}
\end{figure*}

For five-atom linear chains and the combined set in Fig. \ref{Fig15} (a), the accuracy drops below 90\% for the noise level of 7\% (2\%) and significant drops below 80\% for the noise level of 10\% (5\%). On the other hand, the closed configuration (P5) keeps the 100\% accuracy level and 98\%, respectively. Fig. \ref{Fig15} (b) shows the detailed noise dependence. The accuracy of the closed shapes, C4 and P5, is insensitive to the noise due to the distinctive data patterns.

\section{Conclusion}

In summary, we have classified various Rydberg Hamiltonians using two machine learning models. Rydberg identifier (Rydberg-ID) can determine Rydberg Hamiltonians reversely from the time evolution data using machine learning (ML). We compared a support vector machine (SVM) trained with stochastic gradient descent (SGD) and a random forest classifier (RFC). The RFC shows slightly higher accuracy compared to the SVM because the ensemble method (RFC) is more efficient than using a single model (SVM) in our classification of Rydberg atom configurations. We obtain better performance of the RFC, especially for complicated data sets such as the six-atom linear chain. Both models performed exceptionally well in classifying the closed atomic configurations reaching close to 100\% for all the configurations tested.

Both models can classify the number of ground-state atoms in a configuration well (accuracy $> 90\%$). However, the models performed better for the closed shapes with accuracy close to 100\%. We then tested the models' ability to identify the geometric configuration of a fixed number of atoms. The various four atom-connected graphs discussed in Ref. \cite{Kim2020} were classified almost perfectly. [Fig \ref{Fig13}]

Lastly, we assessed the models' performance in classifying all possible Rydberg quantum base states of a fixed atomic configuration. Nonetheless, ML methods converge faster for the closed configuration of C4, P5, and H6 compared to the linear chains for each case. The confusion matrix shows that the distinctive data patterns are related to the high accuracy.

To investigate further the effect on the accuracy, we considered the range of the interaction for the linear chains: NNN vs. NN. Experimentally realistic NNN contributes to better performance in the classification. We also considered combined data sets of closed and linear chains to investigate how the interaction affects the distinction of the probability data's time evolution. Further investigation of machine learning classification for the Rydberg Hamiltonian associated with multi-partite qubit systems paves the way for the diverse usage of ML in Rydberg quantum simulation.  

\begin{acknowledgements} This research was supported by the Air Force Office of Scientific Research (FA2386-20-1-4068), SATU Joint Research Scheme (JRS) (SST004-2020), and University of Malaya Impact Oriented Interdisciplinary Research Grant (IIRG001‐19FNW). JA and MK acknowledge support from Samsung Science and Technology Foundation (SSTF-BA1301-12) and the National Research Foundation of Korea (NRF) (2017R1E1A1A01074307).
\end{acknowledgements}

\bibliography{ML_Ryd_20210728}

\end{document}